\documentclass[a4paper,11pt]{article}
\pdfoutput=1
\usepackage{jheppub}
\usepackage{amssymb}
\usepackage{amsmath}
\usepackage{mathtools}
\usepackage{amsfonts}
\usepackage{dsfont}
\usepackage{young}
\usepackage[vcentermath]{youngtab}
\usepackage{bm}
\usepackage{braket}
\usepackage{simplewick}
\usepackage[offset=1.25em]{simpler-wick}
\usepackage{bm}
\usepackage{subcaption}
\usepackage[most]{tcolorbox}
\usepackage{soul}

\tcbset{colback=yellow!10!white, colframe=red!50!black,
        highlight math style= {enhanced, %<-- needed for the ’remember’ options
            colframe=red,colback=red!10!white,boxsep=0pt}
        }

\usepackage{mathrsfs}
\usepackage{tikz}
\usetikzlibrary{shapes}
\usetikzlibrary{arrows.meta}
\usetikzlibrary{positioning}
\usetikzlibrary{positioning}
\usetikzlibrary{decorations.markings}
\usetikzlibrary{decorations.pathmorphing}

\usepackage{comment}

\numberwithin{equation}{section}

\DeclareMathAlphabet{\mathpzc}{OT1}{pzc}{m}{it}

\newcommand{\mrm}[1]{\mathrm{#1}}
\newcommand{\mcl}[1]{\mathcal{#1}}
\newcommand{\mbb}[1]{\mathbb{#1}}

\newcommand{\mfr}[1]{\mathfrak{#1}}

\newcommand{\wt}[1]{\tilde{#1}}

\newcommand{\be}{\begin{equation}}
\newcommand{\ee}{\end{equation}}

\title{AdS$\mathbf{ _3 \times}$S$\mathbf{^3}$ magnons in the symmetric orbifold}
\author{Matthias R.\ Gaberdiel, Dennis Kempel, and Beat Nairz}
\affiliation{Institut f\"ur Theoretische Physik, ETH Zurich, \\
CH-8093 Z\"urich, Switzerland}
\emailAdd{gaberdiel@itp.phys.ethz.ch, kempeld@student.ethz.ch,
nairzb@student.ethz.ch}

\allowdisplaybreaks[2]

\author{}
\abstract{The ${\rm AdS}_3\times {\rm S}^3$ excitations of string theory on $\text{AdS}_3\times {\rm S}^3\times \mathbb{T}^4$ are identified with certain collective modes in the dual symmetric orbifold. Our identification follows from a careful study of the conformal eigenstates in the perturbed orbifold theory. We find that, in addition to the fractional torus modes (that correspond to the torus excitations in the dual AdS spacetime), there are `long' collective eigenmodes  that involve a superposition of products of fractional torus modes, and that are in natural one-to-one correspondence with the expected ${\rm AdS}_3\times {\rm S}^3$ excitations. These collective modes are deformations of (fractional) ${\cal N}=4$ modes, to which they reduce for integer momentum.}

\emailAdd{}

\begin{document}

\maketitle

\section{Introduction}\label{sec:introduction}

The symmetric orbifold of $\mbb{T}^4$ is the exact CFT dual of string theory on $\text{AdS}_3\times {\rm S}^3\times \mbb{T}^4$ with minimal ($k=1$) NS-NS flux \cite{Gaberdiel:2018rqv,Eberhardt:2018ouy,Eberhardt:2019ywk}. This is a rather special point in moduli space at which the theory exhibits a higher spin symmetry. As a consequence the spectrum is highly degenerate, and, for example, the spacetime interpretation of the individual orbifold states is rather ambiguous.

In order to understand this theory better it is therefore instructive to deform the theory away from this special point. In particular, one can turn on R-R flux in the spacetime, and this corresponds to perturbing the symmetric orbifold theory  by the exactly marginal operator from the $2$-cycle twisted sector, see e.g.\ \cite{David:2002wn}. In general, the corresponding deformation analysis is quite complicated \cite{Lunin:2002fw,Gomis:2002qi,Gava:2002xb,David:2008yk,Pakman:2009zz,Pakman:2009mi,Burrington:2012yq,Gaberdiel:2015uca,Hampton:2018ygz,Guo:2019ady,Lima:2020boh,Guo:2020gxm,Benjamin:2021zkn,Apolo:2022fya,Guo:2022ifr,Hughes:2023fot}, see also \cite{Keller:2019suk}. However, recently \cite{Gaberdiel:2023lco}, two of us showed that the problem becomes tractable in the limit in which one considers states from the $w$-twisted sector with large $w$. In particular, we could give explicit formulae for the anomalous dimensions of essentially all magnon excitations, and the results fit together nicely with the BMN results of  \cite{Berenstein:2002jq}, as well as the predictions from integrability \cite{Babichenko:2009dk,Hoare:2013lja,Borsato:2013qpa,Lloyd:2014bsa,Frolov:2023pjw}.

The analysis of \cite{Gaberdiel:2023lco} concentrated mainly on the fractional torus excitations of the symmetric orbifold which are believed to correspond to the spacetime torus excitations \cite{Frolov:2023pjw}. However, one may wonder whether the ${\rm AdS}_3\times {\rm S}^3$ excitations are also visible in this description. It was proposed in \cite{Lunin:2002fw,Gomis:2002qi,Gava:2002xb} that they should correspond to the fractional ${\cal N}=4$ generators, but since the latter are expressed in terms of bilinears of the torus modes, it is somewhat unclear whether they are in fact independent or not.\footnote{In fact, because of this it was argued in  \cite{Frolov:2023pjw} that there are no such modes present in the symmetric orbifold theory.} In this paper we answer this question in the affirmative: while the identification at the free point is ambiguous, once the theory is perturbed away from the symmetric orbifold point, there is an unambigous way to distinguish between the different types of excitations, and we find that the ${\rm AdS}_3\times {\rm S}^3$ excitations are indeed present in the spectrum. However, while they are closely related to the fractional ${\cal N}=4$ generators as proposed in \cite{Lunin:2002fw,Gomis:2002qi,Gava:2002xb}, this is only true for the global modes; the fractional modes, on the other hand, have a more complicated structure.

In order to arrive at these conclusions we determine (numerically) the eigenstates of the anomalous dimension matrix at large (but finite) $w$. We then study the structure of the different eigenstates as we take $w\rightarrow \infty$. In this limit two distinct classes of eigenstates emerge: the `short' eigenstates which are a sum of a $\mcl{O}(1)$ products of fractional torus modes, and the `long' ones, which involve a linear combination of $\mcl{O}(w)$ such states. The former are just the original torus excitations that were studied in detail in  \cite{Gaberdiel:2023lco}. In this paper we concentrate on the `long' eigenmodes and their associated eigenstates. We show (i) that they are `deformations' of the fractional ${\cal N}=4$ generators (to which they reduce if we consider the global modes); and (ii) that their anomalous dimensions are described by the BMN dispersion relation associated to the massive ${\rm AdS}_3\times {\rm S}^3$ modes \cite{Berenstein:2002jq}. Furthermore the `long' eigenmodes we find are in precise one-to-one correspondence with these ${\rm AdS}_3\times {\rm S}^3$ excitations. Taken together, this is therefore very strong evidence for our identification.
\smallskip

The paper is organised as follows. In Section~\ref{sec:minus}, we briefly review the method of \cite{Gaberdiel:2023lco}, and then describe the eigenstate analysis for one specific bilinear sector (the one where we consider the excitations of two negatively charged fermions). We exhibit the structure of the long eigenstates and calculate their spectrum. While our analysis relies mainly on the large $w$ analysis of \cite{Gaberdiel:2023lco}, we also show that this is legitimate in this context, i.e.\ that the same results are reproduced if one includes the higher magnon corrections of \cite{Gaberdiel:2024nge}. This analysis is then generalised in Section~\ref{sec:others} to other bilinear sectors, and we show there that these `long' eigenstates \emph{only} appear in the sectors in which one should expect them based on the analysis of \cite{Lunin:2002fw,Gomis:2002qi,Gava:2002xb}. Essentially all of these eigenstates are `unphysical' since they do not satisfy the orbifold invariance condition. In Section~\ref{sec:phys} we therefore also analyse physical states in a $3$-magnon sector (made up of three fermionic modes), and reproduce the previous results. Finally, we discuss in Section~\ref{sec:emergent_directions} the connection to the BMN limit, and argue that the long eigenmodes are to be identified with the $\text{AdS}_3\times {\rm S}^3$ excitations in the dual string theory. Section~\ref{sec:concl} contains our conclusions and outlines a number of questions for future work. There are a number of appendices where some of the more technical material is described. We have also included the Jupyter notebook with which our calculations were performed as an ancillary file in the {\tt arXiv} submission.\footnote{The ancillary files also include a Mathematica notebook, based on the code of \cite{Gaberdiel:2024nge}, with which we have confirmed that the approximation of \cite{Gaberdiel:2023lco} is also appropriate here, see for example Section~\ref{sec:unphysical_higher_magnon}.}

\section{The eigenvalue problem in the negative sector}\label{sec:minus}

In this section we review the analysis of \cite{Gaberdiel:2023lco}. We then use it to study the eigenstates of the anomalous dimension matrix at finite $w$, and extrapolate their structure to $w\rightarrow \infty$.

The single particle states in the $w$-cycle twisted sector of the symmetric orbifold are generated by the left-moving modes (we are mostly using the conventions of  \cite{Gaberdiel:2023lco})
\be\label{modes}
\begin{aligned}
& \psi^-( \tfrac{n}{w}) \equiv  \psi^-_{-\frac{1}{2} + \frac{n}{w}} \ , \ \
\psi^+( \tfrac{n}{w}) \equiv \psi^+_{-\frac{3}{2} + \frac{n}{w}} \ ,\qquad
\alpha^i( \tfrac{n}{w}) \equiv \tfrac{1}{\sqrt{ 1-\frac{n}{w}}} \, \alpha^i_{-1 + \frac{n}{w}}  \ \ (i=1,2)  \ ,
\\
& \bar{\psi}^-( \tfrac{n}{w}) \equiv \bar{\psi}^-_{-\frac{1}{2} + \frac{n}{w}} \ , \ \
\bar{\psi}^+( \tfrac{n}{w}) \equiv \bar{\psi}^+_{-\frac{3}{2} + \frac{n}{w}} \ ,\qquad
\bar{\alpha}^i( \tfrac{n}{w}) \equiv \tfrac{1}{\sqrt{1-\frac{n}{w}}} \, \bar{\alpha}^i_{-1 + \frac{n}{w}}  \ \ (i=1,2) \ ,
\end{aligned}
\ee
acting on the upper BPS state, and similarly for the right-movers. The parameter $\frac{n}{w}$ can be thought of as `momentum', $p\equiv \frac{n}{w}$, and the physical states are characterised by the condition that the sum of the left-moving momenta must equal that of the right-moving momenta up to an integer. (This is simply the orbifold invariance condition.) It was argued in \cite{Gaberdiel:2023lco} that, in the large $w$ limit, the system becomes integrable, and that the individual left-moving magnons transform under the  supercharges as
\begin{subequations}\label{commu}
\begin{align}
[Q_1,\alpha^2(p)] &= a(p)\,\psi^-(p)\ , & \{Q_1,\psi^-(p)\} &= 0\ ,\\
[S_1,\alpha^2(p)] &= 0\ , & \{S_1,\psi^-(p)\} &= d(p)\,\alpha^2(p)\ ,\\
[\wt{Q}_2,\alpha^2(p)] &= 0\ ,& \{\wt{Q}_2,\psi^-(p)\} &= b(p)\,\alpha^2(p)\,\mcl{Z}_-\ ,\\
[\wt{S}_2,\alpha^2(p)] &= c(p)\,\psi^-(p)\,\mcl{Z}_+\ , & \{\wt{S}_2,\psi^-(p)\} &= 0 \ ,
\end{align}
\end{subequations}
where
\be\label{N4subset}
Q_1 \equiv G^-_{+\frac{1}{2}} \ , \qquad S_1 \equiv G^+_{-\frac{1}{2}} \ , \qquad
\tilde{Q}_2 \equiv \tilde{G}^{'-}_{+\frac{1}{2}} \ , \qquad \tilde{S}_2 \equiv \tilde{G}^{'+}_{-\frac{1}{2}} \ ,
\ee
and
\begin{equation}\label{eq:coeff_sine_cond}
\bigl( a(p)b(p) \bigr)^* = c(p)d(p) = \frac{g}{2i}\,\bigl(1-e^{-2\pi i p}\bigr)\ .
\end{equation}
The operators $\mcl{Z}_\pm$ change the length of the twist, and the exact momentum conservation in eq.~(\ref{commu}) can strictly only hold for infinite $w$. The description for the other magnon modes (and the remaining supercharges) is similar, see \cite{Gaberdiel:2023lco} for further details. Given that the  supercharges anti-commute to
\be
{\cal C} = \{ Q_1, S_1 \} = (L_0 - K^3_0)  \ , \qquad
\tilde{\cal C}= \{ \tilde{Q}_2, \tilde{S}_2 \}  =  (\tilde{L}_0 - \tilde{K}^3_0)   \ ,
\ee
this then allows one to determine the anomalous conformal dimension $\Delta$ of any state that is generated by these fractional modes. More specifically, it was found that each magnon of `momentum' $p =\frac{n}{w}$ adds to the  total lightcone energy $\Delta={\cal C} + \tilde{\cal C}$ the contribution
\be\label{dispersion}
\epsilon(p) = \sqrt{(1-p)^2 + 4 g^2 \sin^2(\pi p)} = (1-p) + 2 g^2\, \epsilon_2(p) + {\cal O}(g^4)\ ,
\ee
where $g$ is the coupling constant with which the $2$-cycle deformation has been switched on, and
\be\label{omega2}
\epsilon_2(p) = \frac{\sin^2(\pi p)}{(1-p)}
\ee
is the anomalous conformal dimension to second order. On the face of it this would then suggest that the eigenstates with respect to the anomalous dimension are the individual magnon excitations of the form\footnote{For simplicity we shall only consider quarter BPS states here, i.e.\ only apply left-moving magnons.}
\be
A(p_1) \cdots A(p_n) |w\rangle \ ,
\ee
for which the physical state condition becomes $\sum_{i=1}^{n} p_n \in \mathbb{Z}$; to order $g^2$ each such state has then spacetime energy equal to
\be
\Delta = \sum_{i=1}^{n} \epsilon_2(p_i) \ .
\ee
However, as was already noted in \cite{Gaberdiel:2023lco}, the situation must be a bit more subtle: the symmetric orbifold contains for example the ${\cal N}=4$ generators that can be expressed in terms of bilinears of the free fields, and their integer modes are non-anomalous. For example, the $\mathfrak{su}(2)$ $R$-symmetry contains the lowering operator
\be
K^- = -\bar\psi^- \, \psi^- \ ,
\ee
and its zero mode acts on the $w$-cycle twisted sector as
\be\label{Km}
K^-_{0} \, |w\rangle = -\sum_{a=0}^{w} \bar\psi^-\bigl(1-\tfrac{a}{w}\bigr) \, \psi^-\bigl(\tfrac{a}{w}\bigr) \, |w\rangle \ .
\ee
Applied naively, the above formula would then imply that the term in eq.~(\ref{Km}) associated to $a$ has, to second order in perturbation theory, anomalous conformal dimension equal to
\be
\epsilon_2(1-\tfrac{a}{w}) + \epsilon_2\bigl(\tfrac{a}{w}\bigr) \ ,
\ee
where $\epsilon_2$ was defined in (\ref{omega2}).
However, since the ${\cal N}=4$ algebra is not broken by the perturbation, the actual anomalous conformal dimension of the state in (\ref{Km}) must vanish. The resolution of this apparent puzzle was also given in \cite{Gaberdiel:2023lco}: the commutation relations (\ref{commu}), as well as the dispersion relations (\ref{dispersion}) that were derived from it, only hold in the large $w$ limit. At finite $w$, there are correction terms, and we have instead of (\ref{commu})
\be\label{deformedaction}
\begin{array}{llll}
{}[Q_1\, , \alpha^2( \tfrac{n}{w})] & = a^n_n\, \psi^-( \tfrac{n}{w})    \quad
& {}\{Q_1\,, \psi^-( \tfrac{n}{w})\} & = 0 \\[2pt]
{}[S_1\,, \alpha^2( \tfrac{n}{w}) ] & = 0   \quad
& \{S_1\,, \psi^-( \tfrac{n}{w})\} & = d^n_n \, \alpha^2( \tfrac{n}{w})  \\[2pt]
{}[\tilde{Q}_2\, , \alpha^2( \tfrac{n}{w})]  & =  0    \quad
& \{\tilde{Q}_2\,,  \psi^-( \tfrac{n}{w})\}  & =  \sum_m b^m_n \,
\alpha^2( \tfrac{m}{w-1}) \, {\cal Z}_-   \\[2pt]
{} [\tilde{S}_2\,,  \alpha^2( \tfrac{n}{w})]  & =  \sum_m c^m_n \,
\psi^-( \tfrac{m}{w+1}) \, {\cal Z}_+  \quad
& \{\tilde{S}_2\,,  \psi^-( \tfrac{n}{w})\}  & =  0 \ ,
\end{array}
\ee
where the coefficients $b^m_n$ and $c^m_n$ are not just proportional to $\delta_{n,m}$, but rather have a specific form that was worked out there. (The relevant formulae are collected in Appendix~\ref{app:contractions}.) Evaluating the anomalous conformal dimension of (\ref{Km}) at finite $w$ then showed that it is in fact an eigenstate with vanishing anomalous conformal dimension, see \cite[Section~5.7]{Gaberdiel:2023lco}.

While this worked out on the nose, one may be worried about the fact that there is already an implicit large $w$ limit in eq.~(\ref{deformedaction}). Indeed, as was shown in \cite{Gaberdiel:2024nge}, the action of the supercharges also create higher magnon contributions, and even though they are individually supressed in $\frac{1}{w}$, they do modify for example the elements of the mixing matrix even in the large $w$ limit. However, as was also shown in \cite{Gaberdiel:2024nge}, for many quantities of interest, consistently ignoring all of these terms reproduces the correct large $w$ limit.\footnote{For example, including the intermediate higher magnon terms adds higher magnon corrections to the eigenvectors of the mixing matrix. However, the action of the supercharges on these eigenvectors is correctly reproduced by considering only the lowest magnon `heads', and applying the simplified calculation of \cite{Gaberdiel:2023lco} to them.} In the following we shall therefore usually work just with (\ref{deformedaction}), and in most cases this correctly reproduces what one would have obtained if one had included the higher magnon terms (as we have verified explicitly). However, as we explain below, there are a few instances, where these higher order corrections do make a difference for the `long' eigenstates, see in particular the discussion below eq.~(\ref{3.18}); this is not entirely surprising since the definition of the long states is subtle in the large $w$ limit.

\subsection{The full double-magnon spectrum}\label{subsec:minus1}

In this paper we want to study the analogues of the states of the form (\ref{Km}) in more detail. To start with we shall consider the $2$-magnon sector generated by $\bar\psi^- \psi^-$ and allow for `unphysical' states, i.e.\  states whose total momentum  $p=\frac{n}{w}$  is not necessarily an integer.\footnote{We shall come back to the actual physical states in Section~\ref{sec:phys} below.} The relevant space is generated by the $w+1-n$ states of the form
\be\label{eq:minus_space}
\bar\psi^-\bigl(1 +\tfrac{n}{w}- \tfrac{a}{w}\bigr) \, \psi^-\bigl(\tfrac{a}{w}\bigr) \, |w\rangle  \equiv
\bar{\psi}^-_{\frac{1}{2} + \frac{n}{w}-\frac{a}{w}}\psi^-_{-\frac{1}{2}+\frac{a}{w}} \, |w\rangle \ ,
\ee
where $n\leq a \leq w$. The right-moving anomalous conformal dimension operator therefore acts as

\begin{align}
\tilde{\mathcal{C}}\, \bar{\psi}^-_{\frac{1}{2}+\frac{n}{w}-\frac{b}{w}}\psi^-_{-\frac{1}{2}+\frac{b}{w}}\ket{w}
&= \{\tilde{S}_2\,, \tilde{Q}_2\}\, \bar{\psi}^-_{\frac{1}{2}+\frac{n}{w}-\frac{b}{w}}\psi^-_{-\frac{1}{2}+\frac{b}{w}}\ket{w}\nonumber \\
    &= g^2\pi^2 \left(\tfrac{w-1}{w}\right)^2 \sum_{a=n}^w \gamma(\tfrac{n}{w})^a_b\,\, \bar{\psi}^-_{\frac{1}{2}+\frac{n}{w}-\frac{a}{w}}\psi^-_{-\frac{1}{2}+\frac{a}{w}}\ket{w}\ , \label{2.14}
\end{align}
where the matrix elements $\gamma(\tfrac{n}{w})^a_b$ will be determined momentarily. To do so, we first note that $\tilde{S}_2$ annihilates the torus fermions to first order, while $\tilde{Q}_2$ maps the states in (\ref{eq:minus_space}) to states in the $w-1$-cycle twisted sector,
\begin{align}\label{eq:q2_action}
&    \tilde{Q}_2\, \bar{\psi}^-_{\frac{1}{2}+\frac{n}{w}-\frac{b}{w}}\psi^-_{-\frac{1}{2}+\frac{b}{w}}\ket{w}  \\
&  \qquad = g\, \pi\,  \bigl(\tfrac{w-1}{w} \bigr)  \sum_{m=-w+2}^{w-2} \delta^{(w)}(m;n)\sum_{c=m+1}^{w-1}\sqrt{\tfrac{w-1}{c-m}}\, \Big[\kappa_m(\tfrac{n}{w})_b^c \: \alpha^2_{\frac{m-c}{w-1}}\bar{\psi}^-_{-\frac{1}{2}+\frac{c}{w-1}}\nonumber\\
&\hspace{8cm}+\bar{\kappa}_m(\tfrac{n}{w})^c_b\:\bar{\alpha}^2_{\frac{m-c}{w-1}}\psi^-_{-\frac{1}{2}+\frac{c}{w-1}}\Big]\ket{w-1}\ .  \nonumber
\end{align}

\begin{figure}[ht]
\centering
\includegraphics[scale=0.5]{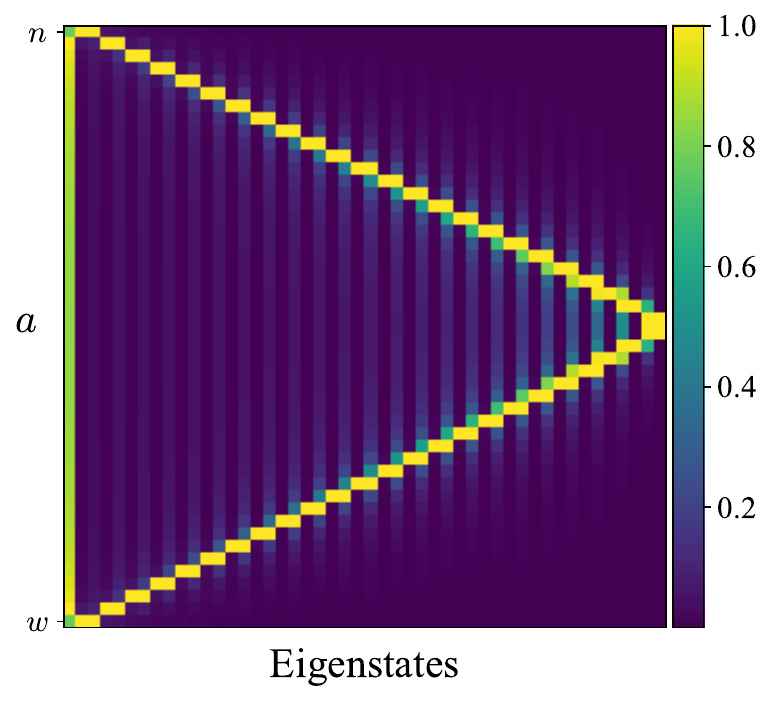}
\caption{The structure of the numerical eigenvalues of the anomalous dimension matrix of eq.~(\ref{2.14}) for the case
$w=50$ and $n=3$. The $x$-axis describes the different eigenstates sorted by their eigenvalue, with the eigenvalue increasing to the right, while the $y$-axis describes the different individual states in (\ref{eq:minus_space}) labelled by $a$ with $n\leq a \leq w$. The color indicates the absolute value of the coefficient of $\bar{\psi}^-_{\frac{1}{2}+\frac{n}{w}-\frac{a}{w}}\psi^-_{-\frac{1}{2}+\frac{a}{w}}\ket{w}$ of the eigenstate, normalised such that the absolute value of the  largest coefficient is equal to 1. The state of minimal eigenvalue is the `long' eigenstate $\Xi^-_{\frac{n}{w}}\ket{w}$ corresponding to the vertical green bar on the left.}
\label{fig:negative_eigenstates}
\end{figure}

Here, we have introduced a  `weighting' factor $\delta^{(w)}(m;n)$, see eq.~(\ref{deltamn}) for its definition, since on the unphysical states  we cannot impose strict momentum conservation at finite $w$. However, since on the actual physical states momentum conservation must be imposed, see \cite{Gaberdiel:2023lco} and Section~\ref{sec:phys} below, we need to keep track of it in this simplified calculation. To do so we have pretended that the state on the left involves an additional `inert' boson with mode number $-\frac{n}{w}$ so that the total state is physical; assuming that this boson is inert, i.e.\ not affected by the action of the supercharges, then allows us to impose strict momentum conservation on the $3$-magnon state (involving the two fermions and the additional boson), and this then gives rise to the `weighting' factor $\delta^{(w)}(m;n)$ using the formulae of \cite{Gaberdiel:2023lco}.\footnote{This factor strongly favours transitions with $\frac{m}{w-1}\approx \frac{n}{w}$. One could alternatively simply set $m=n$ or $m=n-1$, and our results are essentially the same irrespective of which prescription one uses. Obviously, the real justification for this procedure is that the results we find in this section fit nicely together with the physical analysis in the $3$-magnon sector, see Section~\ref{sec:phys}, where this issue does not arise: for physical states we can directly impose momentum conservation.} The coefficients  $\kappa,\bar{\kappa}$ can be calculated from the three-point functions in the perturbed theory, and we express them symbolically as in \cite{Gaberdiel:2023lco}, see Appendix~\ref{app:contractions} for our conventions,

\begin{align}
    \kappa_m(\tfrac{n}{w})^c_b &= \Big(\begin{tikzpicture}[baseline=-.5ex]
    \node[](n1)at(0,0){$\bar{\psi}^+_{r_1}$};
    \node[](n2)[right=-1ex of n1]{$\psi^+_{r_2}$};
    \node[](n3)[right=-1ex of n2]{$\alpha^2_{s_1}$};
    \node[](n4)[right=-1ex of n3]{$\bar{\psi}^-_{s_2}$};
    \draw[](n1.south)-- ([yshift=-1ex]n1.south)--([yshift=-1ex]n3.south)node[midway]{$\times$}--(n3.south);
    \draw[](n2.north)-- ([yshift=1ex]n2.north)--([yshift=1ex]n4.north)--(n4.north);
\end{tikzpicture}\Big)^*\ , & \bar{\kappa}_m(\tfrac{n}{w})^c_b &= - \Big( \begin{tikzpicture}[baseline=-.5ex]
    \node[](n1)at(0,0){$\bar{\psi}^+_{r_1}$};
    \node[](n2)[right=-1ex of n1]{$\psi^+_{r_2}$};
    \node[](n3)[right=-1ex of n2]{$\bar{\alpha}^2_{s_1}$};
    \node[](n4)[right=-1ex of n3]{$\psi^-_{s_2}$};
    \draw[](n1.north)-- ([yshift=1ex]n1.north)--([yshift=1ex]n4.north)--(n4.north);
    \draw[](n2.south)-- ([yshift=-1ex]n2.south)--node[midway]{$\times$}([yshift=-1ex]n3.south)--(n3.south);
\end{tikzpicture}\Big)^*\ , \nonumber
\end{align}
where the mode numbers $r_i,s_i$ are written explicitly in eq.~(\ref{eq:q2_action}).
The action of $\tilde{S}_2$ then maps this state back to a state of the original form, and the matrix element in question, $\gamma(\tfrac{n}{w})^a_b$, is simply
\begin{equation}
    \gamma(\tfrac{n}{w})^a_b = \sum_{m=-w+2}^{w-2} |\delta^{(w)}(m;n)|^2 \sum_{c=m+1}^{w-1}\Big[\kappa^\dagger_m(\tfrac{n}{w})^c_a\:\kappa_m(\tfrac{n}{w})^c_b\:+\:\bar{\kappa}^\dagger_m(\tfrac{n}{w})^c_a\:\bar{\kappa}_m(\tfrac{n}{w})^c_b\Big]\ .
\end{equation}
Using the explicit formulae of \cite{Gaberdiel:2023lco}, we have evaluated the eigenvalues of this `mixing matrix' $\tilde{\mathcal{C}}$ numerically for various values of $w$, and the results are illustrated in
Figure~\ref{fig:negative_eigenstates}.

We observe two qualitatively different classes of eigenstates: `short' eigenstates are composed of only a few $\bar{\psi}^-_{r_1}\psi^-_{r_2} |w\rangle$ states, while `long' eigenstates are a sum of ${\cal O}(w)$ many such states. As we increase $w$, we find that all but one of the eigenstates converge towards a sum (or difference) of two individual short states \eqref{eq:minus_space}, with anomalous dimension
\be\label{eigen1}
\Delta_2 =\epsilon_2(1+\tfrac{n}{w}-\tfrac{a}{w})+\epsilon_2(\tfrac{a}{w}) \ .
\ee
Note that these eigenvalues are two-fold degenerate since the states corresponding to $a$ and $a\mapsto w+n-a$ have
this eigenvalue at large $w$; these eigenstates correspond therefore to the two diagonal yellow lines in Figure~\ref{fig:negative_eigenstates}. The additional long eigenstate, on the other hand, involves all the $w+1-n$ states more or less uniformly; it is described by the first vertical green bar in Figure~\ref{fig:negative_eigenstates}.

\subsection{The structure of the long eigenstate}
\label{subsec:minus2}

\begin{figure}[h]
\centering
\includegraphics[width = \textwidth]{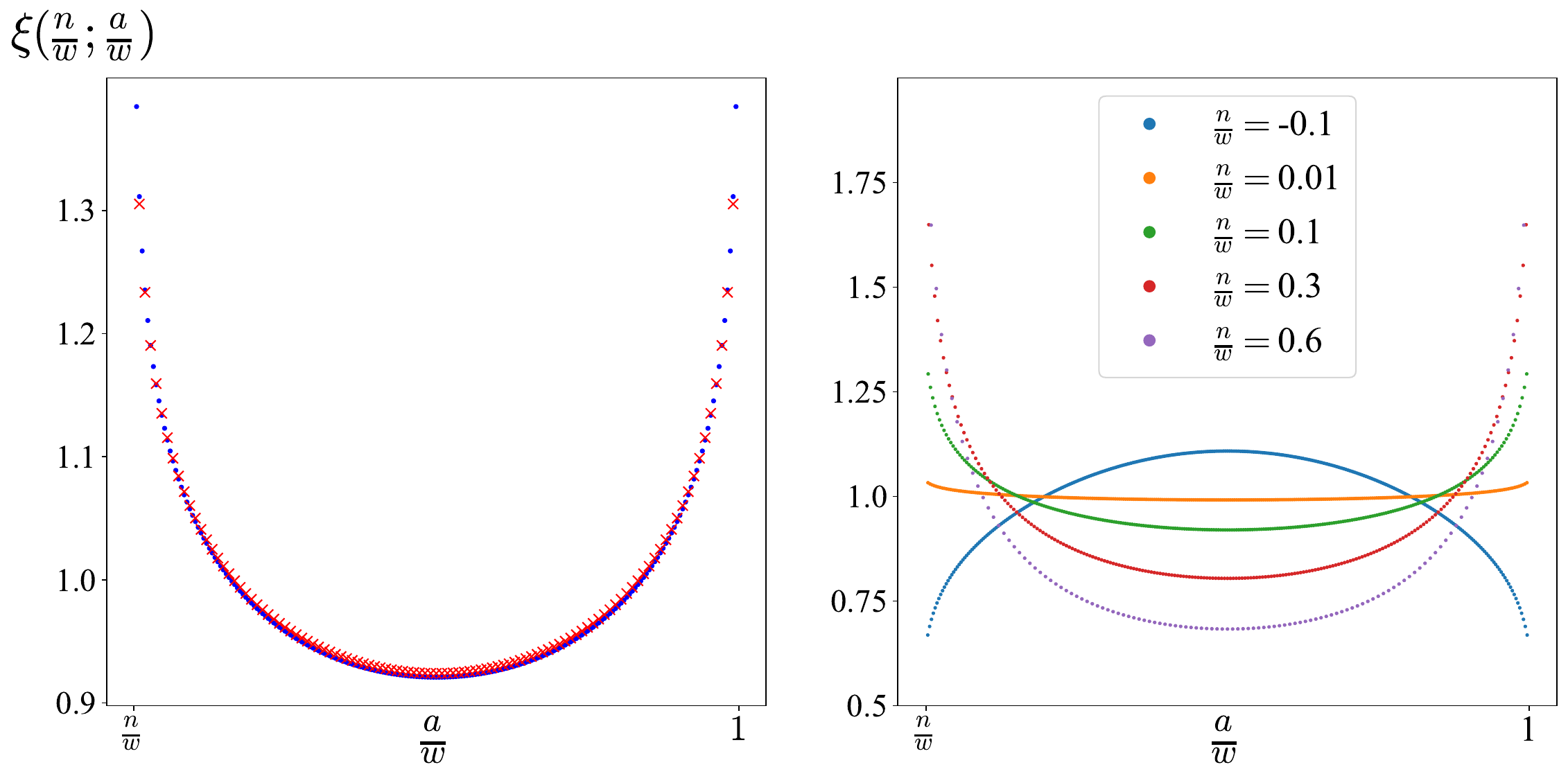}
\caption{Left: The coefficients $\xi(\tfrac{n}{w};\tfrac{a}{w})$ for $\tfrac{n}{w}=0.1$, $w=120$ (red crosses) and $w=240$ (blue dots), plotted against $\tfrac{a}{w}$. Right: The coefficients $\xi(\tfrac{n}{w};\tfrac{a}{w})$ as a function of $\tfrac{a}{w}\in (\tfrac{n}{w},1)$ for $w=300$ and different non-integer values of $\tfrac{n}{w}$.}
\label{fig:Xi_converge}
 \end{figure}

In view of the discussion around eq.~(\ref{Km}) one may naively expect that the long eigenstate is simply the fractional $K^-$ mode,
\be\label{Kmodes}
K^-_{\frac{n}{w}} \, |w\rangle  \equiv -\sum_{a=n}^{w}  \bar{\psi}^-_{\frac{1}{2} + \frac{n}{w}-\frac{a}{w}}\psi^-_{-\frac{1}{2}+\frac{a}{w}} \, |w\rangle \ ,
\ee
but this is not actually correct unless $\frac{n}{w}$ is an integer. Instead it is of the form
\begin{equation}\label{eq:Xi_op}
    \Xi^-_{\frac{n}{w}}\ket{w}= -\tfrac{1}{\sqrt{w+1-n}}\sum_{a=n}^w \xi(\tfrac{n}{w};\tfrac{a}{w})\, \bar{\psi}^-_{\frac{1}{2}+\frac{n}{w}-\frac{a}{w}}\psi^-_{-\frac{1}{2}+\frac{a}{w}}\ket{w},
\end{equation}
where the real coefficients $\xi(\tfrac{n}{w};\tfrac{a}{w})$ depend in general non-trivially on $a$, see Figure~\ref{fig:Xi_converge}.
Note that we have introduced  here the prefactor of $\sqrt{w+1-n}$ so that the $\xi(\tfrac{n}{w};\tfrac{a}{w})$ are typically of $\mcl{O}(1)$ if the state is normalised as
\begin{equation}\label{L2}
    \| \Xi^-_{\frac{n}{w}}\ket{w}\|^2 = \tfrac{1}{w+1-n}\sum_{a=n}^w \xi(\tfrac{n}{w};\tfrac{a}{w})^2 = 1\ .
\end{equation}
For fixed momentum $\tfrac{n}{w}$, the coefficients $\xi(\tfrac{n}{w};\tfrac{a}{w})$ converge towards a continuous function,\footnote{The data are pretty well fitted by the ansatz $\xi(p;q) = a (\sin\pi (1-q)(q-p))^b + c$, where $p\equiv \frac{n}{w}$ and $q\equiv\frac{a}{w}$, and $a$, $b$ and $c$ are certain (rather complicated) functions of $p\equiv \frac{n}{w}$.} see Figure~\ref{fig:Xi_converge}. Note that while the function $\xi(\tfrac{n}{w};\tfrac{a}{w})$ seems to diverge for $\frac{a}{w} \rightarrow \frac{n}{w}$ and $\frac{a}{w} \rightarrow 1$ (for fixed $\frac{n}{w}>0$),\footnote{It appears that the divergence is weaker (or maybe even resolved) if we include the higher magnon corrections of Section~\ref{sec:unphysical_higher_magnon} below. In any case, the value of $\xi$ at the boundaries is a bit ambiguous because the corresponding short eigenstates are degenerate with the long eigenstate, see the discussion at the end of this section. Finally, we note that this divergence is absent for $\frac{n}{w}\leq0$.} the coefficients of the corresponding states (\ref{eq:minus_space}) in (\ref{eq:Xi_op}) tend to zero for large $w$ (because of the square root prefactor). As a consequence the limit state is well-defined. To see that it does not approximate a localised state at large $w$, we have studied its $L^1$ norm defined via
\be\label{eq:L1}
 \|  \Xi^-_{\frac{n}{w}}\ket{w}  \|_{1} = \tfrac{1}{\sqrt{w+1-n}} \sum_{a=n}^{w}\lvert \xi(\tfrac{n}{w};\tfrac{a}{w}) \rvert  \ .
\ee
If we normalise all eigenstates to unity (via the $L^2$ norm, see eq.~(\ref{L2})), the `short' and `long' eigenstates should scale as
\be\label{L1}
\begin{aligned}
&  \qquad \|  \hbox{short eigenstate}  \|_{1}\sim {\cal O}(1) \ ,  \\
&  \qquad \, \, \| \hbox{long eigenstate} \|_{1} \sim {\cal O}(w^{\frac{1}{2}}) \ ,
\end{aligned}
\ee
since for a long eigenstate all coefficients should be proportional to $\xi(\tfrac{n}{w};\tfrac{a}{w})\sim 1$, whereas for a short eigenstate only one coefficient should be non-trivial (and proportional to $w^{\frac{1}{2}}$). If we plot the $L^1$ norm for the normalised eigenstates we can clearly distinguish the long eigenstates from the short ones, see Figure~\ref{fig:2mag_L1} below.

In the first Brillouin zone, i.e.\ for $0<\tfrac{n}{w}<1$, $\xi(\tfrac{n}{w};\tfrac{a}{w})$ is a convex function of $\frac{a}{w}$ with steep slope at the boundary. On the other hand, for small $\tfrac{n}{w}<0$, the function is instead concave in $\frac{a}{w}$, see Figure~\ref{fig:Xi_converge}. Finally, at integer values of $\frac{n}{w}$, the function $\xi(\tfrac{n}{w};\tfrac{a}{w})$ is constant in $\frac{a}{w}$, and the state is just a $K^-$ descendant.

\begin{figure}[h]
\centering
\includegraphics[scale=0.4]{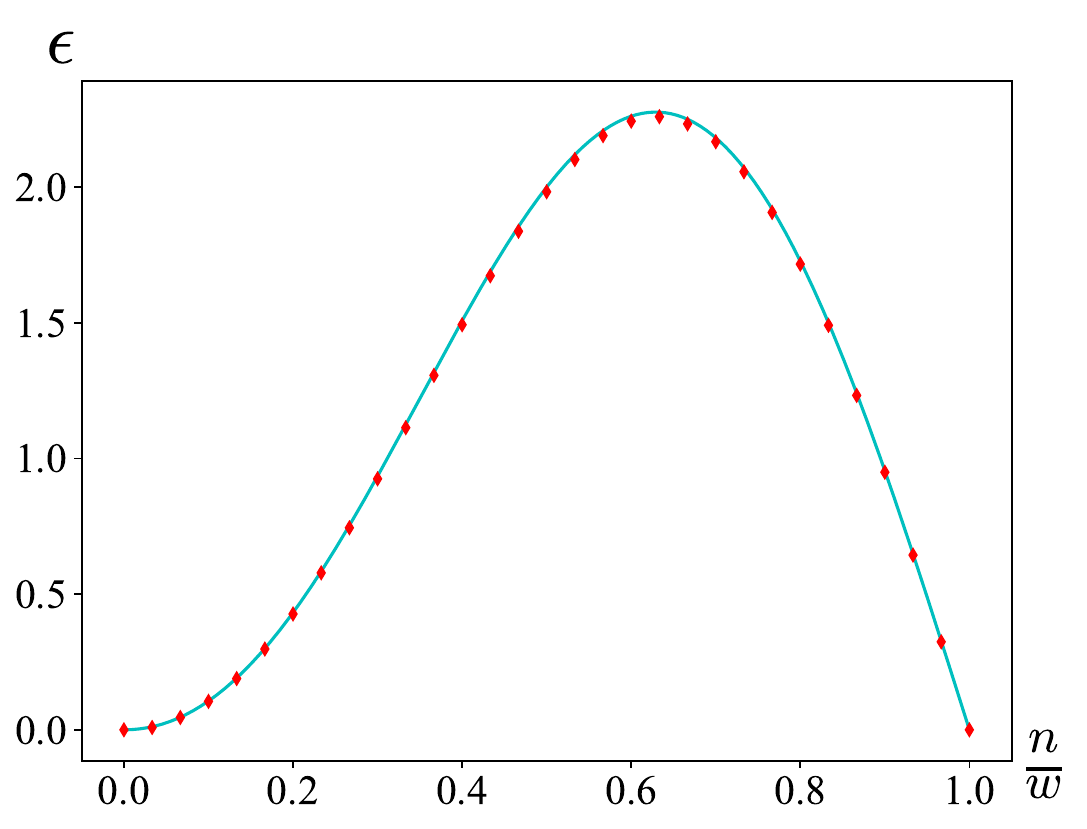}
\caption{The eigenvalues of $\Xi^-_{\frac{n}{w}}\ket{w}$ (red diamonds) and the curve of the function $\epsilon_2(\tfrac{n}{w})$ of eq.~(\ref{omega2}) (cyan) in the first Brillouin zone $\tfrac{n}{w}\in [0,1]$ for $w=300$.}
\label{fig:Xi_disprel}
\end{figure}

We have also worked out the dispersion relation of the long eigenstate, and it approaches, in the large $w$ limit, the same continuous function $\epsilon_2(\tfrac{n}{w})$, see eq.~(\ref{omega2}), as a single torus mode; this can be seen in Figure~\ref{fig:Xi_disprel}. As a consequence the coefficient $\xi(\tfrac{n}{w};\tfrac{a}{w})$ for $a=n$ is ambiguous at infinite $w$ since the long state with $p=\frac{n}{w}$ has then the same eigenvalue as the short state with $\frac{a}{w}=\frac{n}{w}$; indeed, for this value $\Delta_2 = \epsilon_2(\frac{a}{w})$, see eq.~(\ref{eigen1}). The same applies to the coefficient $\xi(\tfrac{n}{w};\tfrac{a}{w})$ for $a=w$. We have therefore always only plotted $\xi(\tfrac{n}{w};\tfrac{a}{w})$ in the range $\frac{n}{w} < \frac{a}{w} < 1$.

\subsection{Higher magnon checks}\label{sec:unphysical_higher_magnon}

\begin{figure}[ht]
\centering
\includegraphics[scale = 0.4]{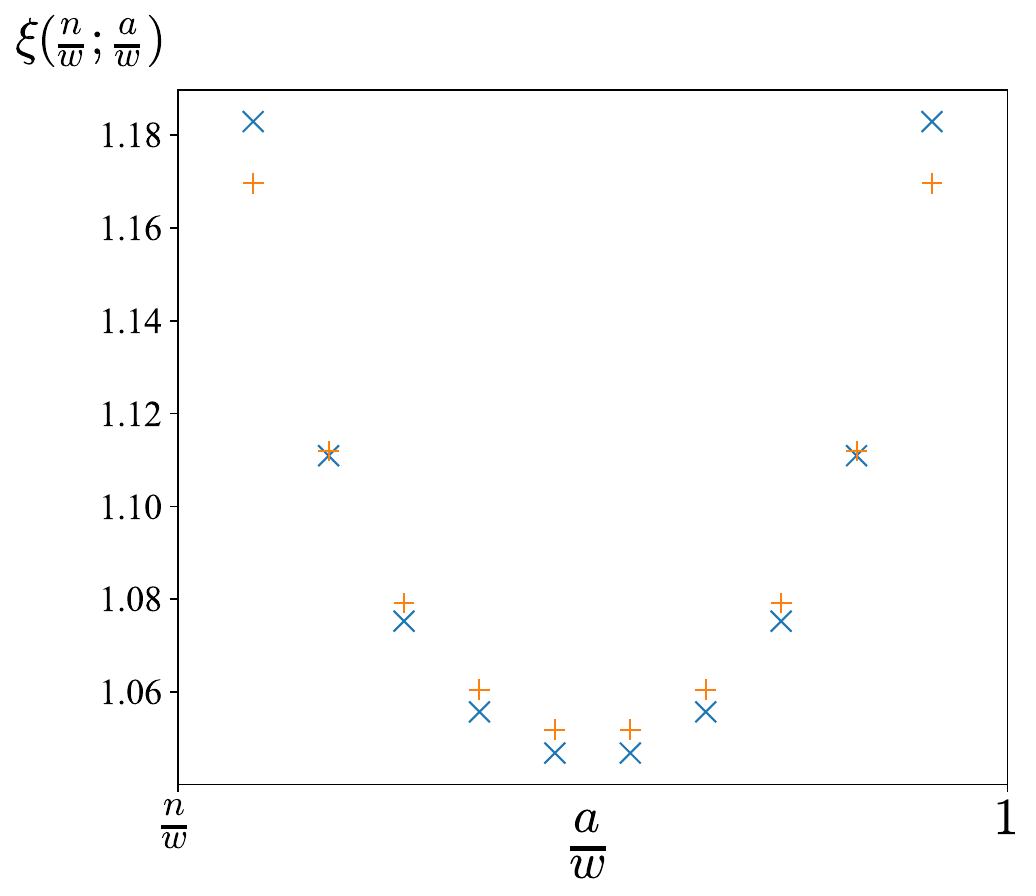}
\caption{The long eigenstate for $w=12$, $n=1$, when only taking into account two-magnon transitions (blue crosses) and when including four-magnon states (orange pluses).}\label{fig:unphysical_comparison}
\end{figure}

In the above analysis we have only considered, as in \cite{Gaberdiel:2023lco}, intermediate $2$-magnon states. For the evaluation of the anomalous conformal dimensions of the $2$-magnon states, this approximation is justified \cite{Gaberdiel:2024nge}. However, since we are now also considering the `long' states that go beyond the naive large $w$ limit, we need to check that this continues to be the case. We have therefore also applied the technology of \cite{Gaberdiel:2024nge} to our situation, and included transitions to states with more modes. By considering the eigenstates that have the largest overlap with the two-magnon states, we again find that there is one long eigenstate. With the inclusion of four-magnon states, we have checked this up to $w=12$ and $n=1$ (for these values, 892 states can mix instead of the 12 states when restricting to two-magnon transitions). The result is shown in Figure~\ref{fig:unphysical_comparison}.

\section{Other sectors}
\label{sec:others}

In the previous section we have analysed the eigenstates (and eigenvalues) in the sector generated by $\bar\psi^- \psi^-$, but it is instructive to repeat this exercise also for the other bilinear sectors. The situation is essentially the same among the bilinear eigenstates of the form
\be
G^- \sim \bar{\alpha}^1 \psi^- + \alpha^1 \bar{\psi}^- \ ,   \qquad
G^{'-} \sim  \bar{\alpha}^2 \psi^- + \alpha^2 \bar{\psi}^- \ ,
\ee
or states in the neutral sector generated by the neutral bilinears
\be
L \ , \ \ K^3 \ \sim \ \ \bar{\alpha}^1 \alpha^2 - \bar{\alpha}^2 \alpha^1 \ , \ \ \bar\psi^+ \psi^- + \bar\psi^- \psi^+ \ .
\ee
However, the `positive' sectors, i.e.\ the states generated by the bilinear terms of the form
\be
K^+ \sim \bar\psi^+ \psi^+ \ , \ \ G^+ \sim \bar{\alpha}^2 \psi^+ + \alpha^2 \bar{\psi}^+ \ ,   \qquad
G^{'+} \sim  \bar{\alpha}^1 \psi^+ + \alpha^1 \bar{\psi}^+ \ ,
\ee
behave differently in that no `long' eigenstate emerges. Since this is quite important for the interpretation in terms of ${\rm AdS}_3 \times {\rm S}^3$ modes, let us give some details for the example of the $\bar\psi^+ \psi^+$ sector. In Section~\ref{subsec:others2} we shall then also describe briefly the results in the $G^-$, $G'^-$, and neutral sectors, and discuss some subtleties that arise there.

\subsection{The positive sector}\label{subsec:others1}

\begin{figure}[h]
\centering
\includegraphics[scale=0.6]{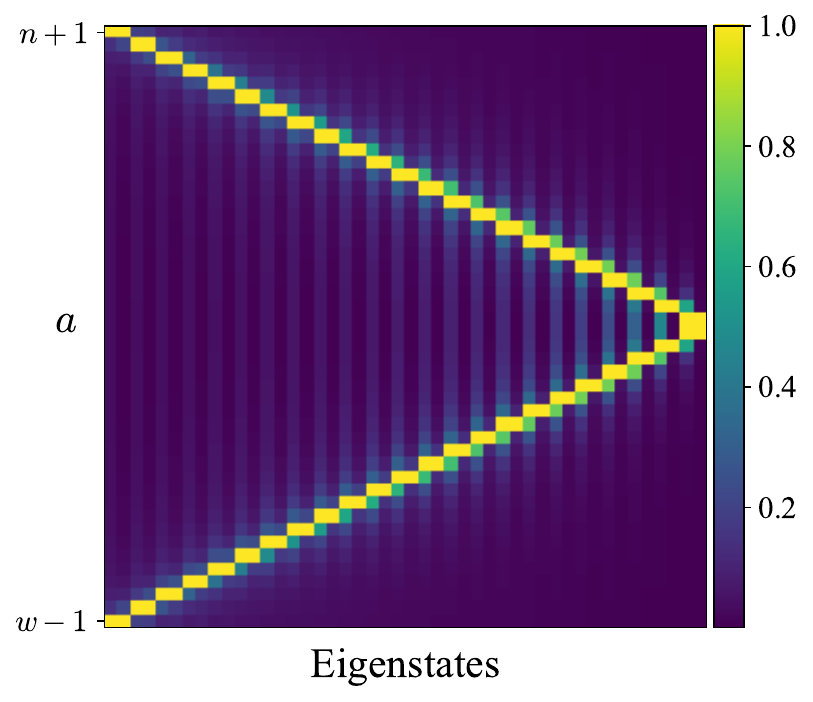}
\caption{The structure of the numerical eigenvalues of the anomalous dimension matrix of eq.~(\ref{3.6})  for $w=50$ and $n=3$. Unlike the negative sector in Figure~\ref{fig:negative_eigenstates}, there is no long eigenstate, signalled by a vertical green bar, here.}
\label{fig:positive_eigenstates}
\end{figure}

The analysis for the positive sector of $K^+$ is technically very similar to that of Section~\ref{subsec:minus1}. Instead of (\ref{eq:minus_space}), the space is now spanned by\footnote{The corresponding global ${\cal N}=4$ modes are the $K^+_n$ modes that vanish unless $n\leq -1$. In this case, their action also involves fermions acting on the remaining $N-w$ copies, i.e.\ $K^+_{-n}$ with $n\in \mathbb{N}$ does not map single particle states to single particle states.}
\begin{equation}\label{eq:plus_space}
    \bar{\psi}^+_{-\frac{1}{2}+\frac{n}{w}-\frac{a}{w}}\psi^+_{-\frac{3}{2}+\frac{a}{w}}\ket{w}\ ,
\end{equation}
where $n+1\leq a \leq w-1$, and the uniform linear combination now agrees with $K^+_{-2+n/w}\ket{w}$. These states are annihilated by the supercharge $\tilde{Q}_2$, while $\tilde{S}_2$ maps them onto intermediate states of cycle length $w+1$,
\begin{align}\label{eq:s2_action}
&    \tilde{S}_2\, \bar{\psi}^+_{-\frac{1}{2}+\frac{n}{w}-\frac{b}{w}}\psi^+_{-\frac{3}{2}+\frac{b}{w}}\ket{w} \\
& \qquad = g\pi \bigl(\tfrac{w}{w+1}\bigr)\, \sum_{m=-w}^{w}\hat\delta^{(w)}(m;n)\sum_{c=m+1}^{w}\sqrt{\tfrac{w+1}{c-m}}\Big[\hat{\kappa}_m(\tfrac{n}{w})_b^c \: \alpha^1_{\frac{m-c}{w+1}}\bar{\psi}^+_{-\frac{3}{2}+\frac{c}{w+1}}\nonumber\\
&\hspace{8cm}  +\hat{\bar{\kappa}}_m(\tfrac{n}{w})^c_b\:\bar{\alpha}^1_{\frac{m-c}{w+1}}\psi^+_{-\frac{3}{2}+\frac{c}{w+1}}\Big]\ket{w+1}\ \nonumber,
\end{align}
where the weighting function $\hat{\delta}^{(w)}(m;n)$ has a similar origin as before, see eq.~(\ref{deltahatmn}) for the precise definition, and the transition amplitudes are
\begin{align}
    \hat{\kappa}_m(\tfrac{n}{w})^c_b = -\begin{tikzpicture}[baseline=-.5ex]
    \node[](n1)at(0,0){$\psi^-_{s_1}$};
    \node[](n2)[right=-1ex of n1]{$\bar{\alpha}^2_{s_2}$};
    \node[](n3)[right=-1ex of n2]{$\bar{\psi}^+_{r_1}$};
    \node[](n4)[right=-1ex of n3]{$\psi^+_{r_2}$};
    \draw[](n1.north)-- ([yshift=1ex]n1.north)--([yshift=1ex]n3.north)--(n3.north);
    \draw[](n2.south)-- ([yshift=-1ex]n2.south)--node[midway]{$\times$}([yshift=-1ex]n4.south)--(n4.south);
\end{tikzpicture}\ ,\qquad
    \hat{\bar{\kappa}}_m(\tfrac{n}{w})^c_b = \begin{tikzpicture}[baseline=-.5ex]
    \node[](n1)at(0,0){$\bar{\psi}^-_{s_1}$};
    \node[](n2)[right=-1ex of n1]{$\alpha^2_{s_2}$};
    \node[](n3)[right=-1ex of n2]{$\bar{\psi}^+_{r_1}$};
    \node[](n4)[right=-1ex of n3]{$\psi^+_{r_2}$};
    \draw[](n1.north)-- ([yshift=1ex]n1.north)--([yshift=1ex]n4.north)--(n4.north);
    \draw[](n2.south)-- ([yshift=-1ex]n2.south)--node[midway]{$\times$}([yshift=-1ex]n3.south)--(n3.south);
\end{tikzpicture}\ .
\end{align}
Again, the action of $\tilde{Q}_2$ then maps these states back to the states of the original form, and we find
\begin{align}
\tilde{\mathcal{C}}\, \bar{\psi}^+_{-\frac{1}{2}+\frac{n}{w}-\frac{b}{w}}\psi^+_{-\frac{3}{2}+\frac{b}{w}}\ket{w}
&= g^2\pi^2 \left(\tfrac{w}{w+1}\right)^2\sum_{a=n+1}^{w-1} \hat{\gamma}(\tfrac{n}{w})^a_b\,\, \bar{\psi}^+_{-\frac{1}{2}+\frac{n}{w}-\frac{a}{w}}\psi^+_{-\frac{3}{2}+\frac{a}{w}}\ket{w}\ , \label{3.6}
\end{align}
where now
\begin{equation}
    \hat{\gamma}(\tfrac{n}{w})^a_b = \sum_{m=-w}^{w}|\hat\delta^{(w)}(m;n)|^2\sum_{c=m+1}^{w}\Big[\hat\kappa_m^\dagger(\tfrac{n}{w})^c_a\:\hat\kappa_m(\tfrac{n}{w})^c_b\:+\:\hat{\bar{\kappa}}_m^\dagger(\tfrac{n}{w})^c_a\: \hat{\bar{\kappa}}_m(\tfrac{n}{w})^c_b\Big]\ .
\end{equation}
Upon diagonalising the mixing matrix  $\hat{\gamma}(\tfrac{n}{w})^a_b$, the spectrum of anomalous dimensions can be determined at finite $w$, and the structure of the eigenstates is shown in Figure~\ref{fig:positive_eigenstates}.

\begin{figure}
\centering
\includegraphics[scale=0.4]{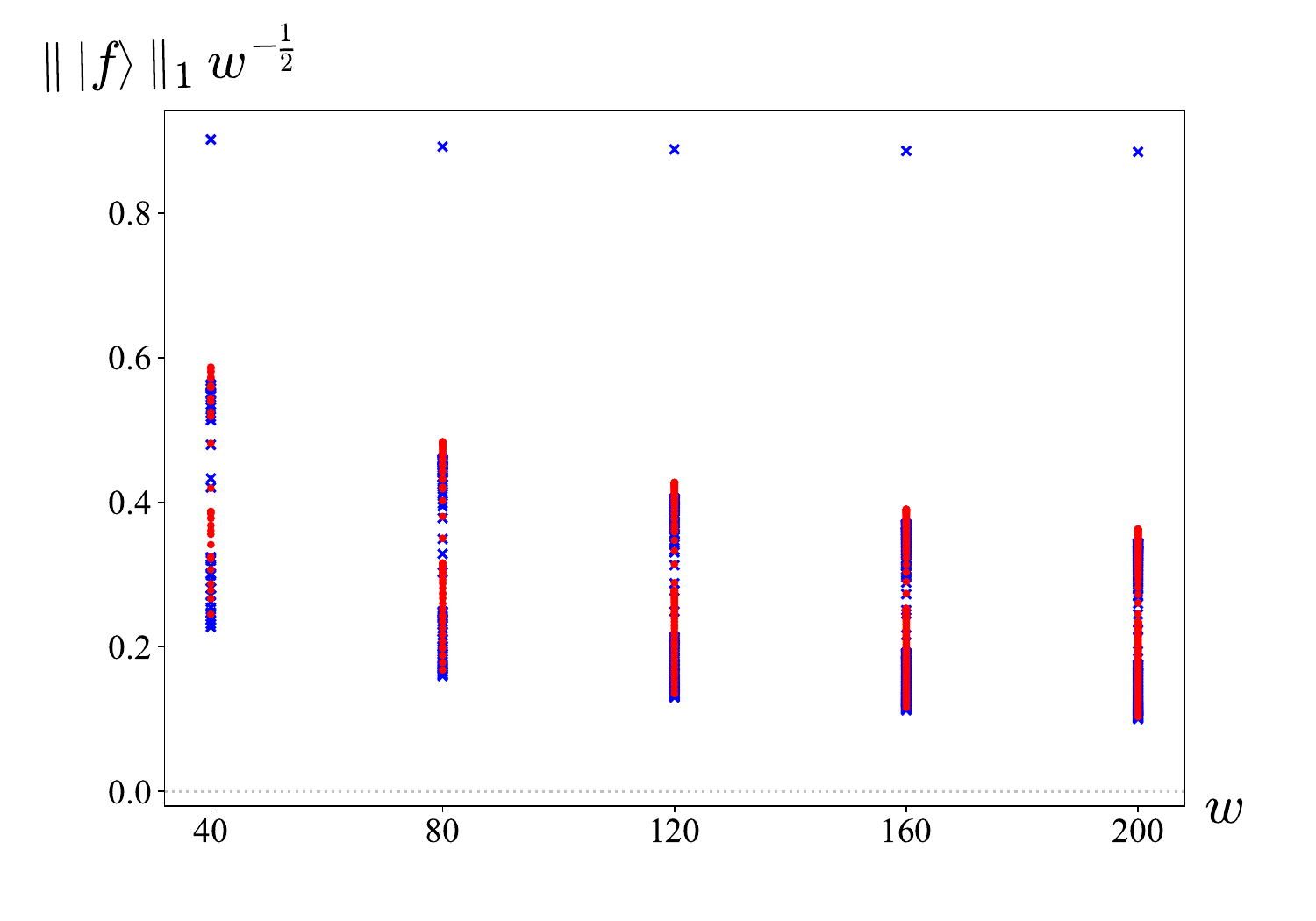}
\caption{The $L^1$ norm \eqref{eq:L1} times the factor $w^{-\frac{1}{2}}$ of all eigenstates of the positive (red) and negative (blue) two magnon systems as a function of the cycle length $w$ for fixed momentum $\frac{n}{w}=0.2$. The negative sector, i.e.\ the states spanned by (\ref{eq:minus_space}), contains one long eigenstate whose $L^1$ norm grows as $w^{\frac{1}{2}}$, which appears at the top of the figure. On the other hand, in the positive sector, i.e. among the states spanned by (\ref{eq:plus_space}), there is no such state.}
\label{fig:2mag_L1}
\end{figure}

The main new feature relative to the discussion of Section~\ref{subsec:minus1} is that now there is no long eigenstate. In order to see this more quantitatively, we have evaluated the $L^1$ norm of these states, following the discussion around eq.~(\ref{eq:L1}). The scaling behaviour of the eigenstates in the negative and positive sectors, i.e.\ for the states that are linear combinations of (\ref{eq:minus_space}) and (\ref{eq:plus_space}), respectively, is shown in Figure~\ref{fig:2mag_L1}.

%If we normalise all eigenstates to
%Using the $L^1$ norm One way to make this distinction more quantitative is to introduce the $L^1$ norm among the relevant states via
%\be\label{eq:L1}
%\ket{f}=\sum_{a=n+1}^{w-1} f(\tfrac{n}{w};\tfrac{a}{w})\bar{\psi}^+_{-\frac{1}{2}+\frac{n}{w}-\frac{a}{w}}\, \psi^+_{-\frac{3}{2}+\frac{a}{w}}\ket{w} \ , \qquad   \| \ket{f} \|_{1} = \sum_{a=n+1}^{w-1}\lvert f(\tfrac{n}{w};\tfrac{a}{w}) \rvert \ ,
%\ee
%and similarly for the states generated by (\ref{eq:minus_space}). If we normalise all eigenstates to unity (via the $L^2$ norm, see eq.~(\ref{L2})), then `short' and `long' eigenstates should scale
%\be\label{L1}
%\begin{aligned}
%& \hbox{short eigenstate:} \qquad \| \ket{f} \|_{1} \sim {\cal O}(1) \ ,  \\
%& \hbox{long eigenstate:} \qquad \, \, \| \ket{f} \|_{1} \sim {\cal O}(w^{\frac{1}{2}}) \ ,
%\end{aligned}
%\ee
%since for a long eigenstate all coefficients should be proportional to $f(\tfrac{n}{w};\tfrac{a}{w})\sim w^{-\frac{1}{2}}$.  The scaling behaviour of the eigenstates in the negative and positive sectors, i.e.\ for the states in (\ref{eq:minus_space}) and (\ref{eq:plus_space}), respectively, is shown in Figure~\ref{fig:2mag_L1}.

One may wonder what selects the positive versus the negative sector, and it is clear that this is a consequence of our convention for the BPS state, i.e.\ that we choose the `chiral primary' state in the BPS multiplet (with eigenvalue $K^3_0=h=j=\frac{w+1}{2}$), rather than the anti-chiral primary with eigenvalue $K^3_0=-\frac{w+1}{2}$.\footnote{On the other hand, we have checked that the results are unaltered if we consider instead of the `top' BPS state with $h=j=\frac{w+1}{2}$, the `bottom' BPS state, see eq.~(\ref{bottomBPS}), with $h=j=\frac{w-1}{2}$ or either of the two `middle' BPS states with $h=j=\frac{w}{2}$, as long as we always work with the chiral primary state whose $K^3_0$ eigenvalue is $K^3_0=j$.}

\subsection{Other long eigenstates}\label{subsec:others2}

The analysis in the other sectors is essentially the same, so let us close this section by just summarising the long eigenstates we found: in addition to the state $\Xi^-_{\frac{n}{w}}\ket{w}$ from eq.~(\ref{eq:Xi_op}), they are
\begin{align}\label{eq:other_bilinear_magnons}
    \Gamma^-_{-\frac{1}{2}+\frac{n}{w}}\ket{w} &= \tfrac{1}{\sqrt{(w+1-n)(1-\frac{n}{w})}}\sum_{a=n+1}^w\xi(\tfrac{n}{w};\tfrac{a}{w})\Big[\alpha^1_{\frac{n}{w}-\frac{a}{w}}\bar{\psi}^-_{-\frac{1}{2}+\frac{a}{w}}
    +\bar{\alpha}^1_{\frac{n}{w}-\frac{a}{w}}\psi^-_{-\frac{1}{2}+\frac{a}{w}}\Big]\ket{w}\ , \nonumber \\
    \nonumber \\
    \Gamma'^-_{-\frac{1}{2}+\frac{n}{w}}\ket{w} &= \tfrac{1}{\sqrt{(w+1-n)(1-\frac{n}{w})}}\sum_{a=n+1}^w\xi(\tfrac{n}{w};\tfrac{a}{w})\Big[\alpha^2_{\frac{n}{w}-\frac{a}{w}}\bar{\psi}^-_{-\frac{1}{2}+\frac{a}{w}}
    +\bar{\alpha}^2_{\frac{n}{w}-\frac{a}{w}}\psi^-_{-\frac{1}{2}+\frac{a}{w}}\Big]\ket{w}\ , \nonumber\\
    \nonumber \\
    \Lambda_{-1+\frac{n}{w}}\ket{w} &= \tfrac{1}{\sqrt{(w+1-n)(1-\frac{n}{w})^2}}\Bigg(\sum_{a=n+1}^{w-1}\xi(\tfrac{n}{w};\tfrac{a}{w})\:\Big[\bar{\alpha}^1_{\frac{n}{w}-\frac{a}{w}}\alpha^2_{-1+\frac{a}{w}}
    -\alpha^1_{\frac{n}{w}-\frac{a}{w}}\bar{\alpha}^2_{-1+\frac{a}{w}}\Big] \nonumber \\
    &\quad +\sum_{a=n+1}^w \xi(\tfrac{n}{w};\tfrac{a}{w}) \: (\tfrac{a}{w}-\tfrac{n}{w})\Big[\bar{\psi}^+_{-\frac{1}{2}+\frac{n}{w}-\frac{a}{w}}\psi^-_{-\frac{1}{2}+\frac{a}{w}}
    - \psi^+_{-\frac{1}{2}+\frac{n}{w}-\frac{a}{w}}\bar{\psi}^-_{-\frac{1}{2}+\frac{a}{w}}\Big]\Bigg)\ket{w}\ ,
\end{align}
where in each case the function $\xi(\tfrac{n}{w};\tfrac{a}{w})$ is \emph{the same}, i.e.\ the one sketched in Figure~\ref{fig:Xi_converge}. If $\frac{n}{w}$ is an integer, $\xi(\tfrac{n}{w};\tfrac{a}{w})$ becomes constant, and these states are simply the descendant states of
\be
G^-_{-\frac{1}{2}+\frac{n}{w}}\ , \qquad G'^-_{-\frac{1}{2}+\frac{n}{w}} \ , \qquad \hbox{and} \qquad
(L+\partial K^3)_{-1+\frac{n}{w}}=L_{-1+\frac{n}{w}}-\tfrac{n}{w}K^3_{-1+\frac{n}{w}} \ ,
\ee
respectively.  On the other hand, the positive generators $K^+_{-2}$, $G^+_{-3/2}$ and $G'^+_{-3/2}$ do not give rise to similar eigenstates for arbitrary (fractional) momenta. We note that the long eigenstates above are mapped into one another by the action of the left-moving supercharges,
\begin{align}
    G'^+_{-\frac{1}{2}}\Xi^-_{\frac{n}{w}}\ket{w} &= -\sqrt{1-\tfrac{n}{w}}\, \Gamma^-_{-\frac{1}{2}+\frac{n}{w}}\ket{w} \ , \\
     G^+_{-\frac{1}{2}}\Xi^-_{\frac{n}{w}}\ket{w} & = \sqrt{1-\tfrac{n}{w}}\, \Gamma'^-_{-\frac{1}{2}+\frac{n}{w}}\ket{w}, \\
    G^+_{-\frac{1}{2}}\Gamma^-_{-\frac{1}{2}+\frac{n}{w}}\ket{w} &= G'^+_{-\frac{1}{2}} \Gamma'^-_{-\frac{1}{2}+\frac{n}{w}}\ket{w} = \sqrt{1-\tfrac{n}{w}}\, \Lambda_{-1+\frac{n}{w}}\ket{w},
\end{align}
which follows directly from the (anti-)commutation relations of the torus modes. The normalisation of the states also follows directly from the $\mathcal{N}=4$ algebra, since, for example,
\begin{equation}
    \|\Gamma^-_{-\frac{1}{2}+\frac{n}{w}}\ket{w}\|^2 =  \tfrac{1}{1-\frac{n}{w}}\bra{w}(\Xi^-_{\frac{n}{w}})^\dagger(L_0-K^3_0)\,  \Xi^-_{\frac{n}{w}}\ket{w} = 1 + {\cal O}(g^2)\ ,
\end{equation}
where we have used that
\be\label{3.18}
\{G'^-_{+\frac{1}{2}},G'^+_{-\frac{1}{2}}\}=L_0-K_0^3 \ , \quad \hbox{and} \quad
(L_0-K^3_0)\, \Xi^-_{\frac{n}{w}}\ket{w} = (1-\tfrac{n}{w} + {\cal O}(g^2))\, \Xi^-_{\frac{n}{w}}\ket{w}\ .
\ee
Furthermore, all of these states have the same dispersion relation as that of a single torus magnon,
eq.~(\ref{omega2}), see Figure~\ref{fig:Xi_disprel}.

However, unlike the situation in the negative sector above, see Section~\ref{sec:unphysical_higher_magnon}, for some of these calculations the inclusion of higher magnon modes is important. In particular, the anomalous dimensions are not equal when calculating them with either the anti-commutator $\{\tilde{S}_1,\tilde{Q}_1\}$ or $\{\tilde{S}_2,\tilde{Q}_2\}$, but there is an element-wise $\mcl{O}\big(\frac{1}{w}\big)$ difference. While these small deviations do not affect the short eigenstates, they can affect the long ones, for which $\mcl{O}\big(\frac{1}{w}\big)$ terms are by construction relevant, see the discussion in Section \ref{sec:minus}. Indeed, depending on the anti-commutator one works with, one observes an \emph{additional} long state in the $G^-$, $G'^-$, and $L$ sectors. Again using the techniques (and in particular the Mathematica code) of \cite{Gaberdiel:2024nge} we have checked that this extra state disappears once the higher magnon terms are included, i.e.\ that it is an artefact of cumulative $\mcl{O}\big(\frac{1}{w}\big)$ errors. Obviously, the full calculation of \cite{Gaberdiel:2024nge} is much more costly, and one cannot push the analysis to very large values of $w$. We have noted that a `practical' way of sidestepping this problem is to always work with the sum of the two anti-commutators; we do not really understand why this prescription works, but we have checked that it does, i.e.\ it always produces the correct eigenstates (as confirmed by the full analysis of \cite{Gaberdiel:2024nge}), and this analysis can then be pushed easily to large values of $w$.

\section{Physical states}\label{sec:phys}

Up to now we have considered mainly unphysical states since we have not imposed the orbifold invariance (or integer momentum) condition. One may therefore be worried that the above results are an artefact of this simplifying assumption. In this section we look at physical states in the sector involving two $\psi^-$ modes and one $\bar{\psi}^-$ mode, and repeat the above analysis. As we will see, the `long' eigenstates from above will also appear in this description, thus showing that they are a true feature of the symmetric orbifold. We will also briefly describe the corresponding positive sector (consisting of two $\psi^+$ modes and one $\bar{\psi}^+$ mode) and also see that it behaves analogously to the unphysical calculation, i.e.\ that it does not exhibit any signs of a `long' eigenstate. We have also repeated the analysis for
all the other three-magnon sectors, and in each case the long eigenstates we find correspond to the states in  eq.~(\ref{eq:Xi_op}) or eq.~(\ref{eq:other_bilinear_magnons}), respectively.

\subsection{The negative three-fermion sector}

Let us consider the states of the form
\begin{equation}\label{eq:3mag_neg_basis}
    \bar{\psi}^-_{\frac{1}{2}+\frac{n}{w}-\frac{a}{w}}\psi^-_{-\frac{1}{2}+\frac{a}{w}}\psi^-_{-\frac{1}{2}+k-\frac{n}{w}}\, |w\rangle\ ,
\end{equation}
with $n\leq a \leq w$ and net momentum $k\in \mathbb{Z}$ with $k\leq 1$. The internal momentum $\tfrac{n}{w}$ ranges then over the first $2-k$ Brillouin zones, $w(k-1) \leq n \leq w$; in the following we shall therefore restrict our attention to $k=1$. Furthermore, due to the presence of two $\psi^-$ modes, the states labelled by $(n,a)$ and $(w-a,w-n)$ are identical. The dimension of this space of states then scales as $w^2$.

Paralleling our previous analysis, we now study the action of the supercharge $\tilde{Q}_2$ on these basis states,
\begin{align}\label{eq:q2_action_physical}
\tilde{Q}_2\, \bar{\psi}^-_{\frac{1}{2}+\frac{n}{w}-\frac{b}{w}}\psi^-_{-\frac{1}{2}+\frac{b}{w}}\psi^-_{\frac{1}{2}-\frac{n}{w}}\ket{w}&
 \nonumber \\
 =  g\pi \bigl(\tfrac{w-1}{w}\bigr)\, \times   \Bigg[\sum_{l=0}^{w-2}\: \sum_{c=0}^{w-2-l}&\bigl(1-\tfrac{l+c}{w-1}\bigr)^{-\frac{1}{2}}\, \:\kappa_{n,b}^{l,c} \: \alpha^2_{-1+\frac{l+c}{w-1}}\bar{\psi}^-_{\frac{1}{2}-\frac{c}{w-1}}\psi^-_{\frac{1}{2}-\frac{l}{w-1}} \\
     +\sum_{l=0}^{\lfloor \frac{w-3}{2}\rfloor}\:\sum_{c=l+1}^{w-2-l}&\bigl(1-\tfrac{l+c}{w-1}\bigr)^{-\frac{1}{2}}\, \:\bar{\kappa}_{n,b}^{l,c}\:\bar{\alpha}^2_{-1+\frac{l+c}{w-1}}\psi^-_{\frac{1}{2}-\frac{c}{w-1}}\psi^-_{\frac{1}{2}-\frac{l}{w-1}}\Bigg]\ket{w-1}\ , \nonumber
\end{align}
where
\begin{align}
    \kappa_{n,b}^{l,c} = \Big(\begin{tikzpicture}[baseline=-.5ex]
    \node[](n1)at(0,0){$\bar{\psi}^+_{r_1}$};
    \node[](n2)[right=-1ex of n1]{$\bar{\psi}^+_{r_2}$};
    \node[](n3)[right=-1ex of n2]{$\psi^+_{r_3}$};
    \node[](n4)[right=-1ex of n3]{$\alpha^2_{s_1}$};
    \node[](n5)[right=-1ex of n4]{$\bar{\psi}^-_{s_2}$};
    \node[](n6)[right=-1ex of n5]{$\psi^-_{s_3}$};
    \draw[](n2.south)-- ([yshift=-1ex]n2.south)--([yshift=-1ex]n4.south)node[midway]{$\times$}--(n4.south);
    \draw[](n1.north)-- ([yshift=1ex]n1.north)--([yshift=1ex]n6.north)--(n6.north);
    \draw[](n3.north)-- ([yshift=2.5ex]n3.north)--([yshift=2.5ex]n5.north)--(n5.north);
\end{tikzpicture} - \begin{tikzpicture}[baseline=-.5ex]
    \node[](n1)at(0,0){$\bar{\psi}^+_{r_1}$};
    \node[](n2)[right=-1ex of n1]{$\bar{\psi}^+_{r_2}$};
    \node[](n3)[right=-1ex of n2]{$\psi^+_{r_3}$};
    \node[](n4)[right=-1ex of n3]{$\alpha^2_{s_1}$};
    \node[](n5)[right=-1ex of n4]{$\bar{\psi}^-_{s_2}$};
    \node[](n6)[right=-1ex of n5]{$\psi^-_{s_3}$};
    \draw[](n1.south)-- ([yshift=-1ex]n1.south)--([yshift=-1ex]n4.south)node[midway]{$\times$}--(n4.south);
    \draw[](n2.north)-- ([yshift=1ex]n2.north)--([yshift=1ex]n6.north)--(n6.north);
    \draw[](n3.north)-- ([yshift=2.5ex]n3.north)--([yshift=2.5ex]n5.north)--(n5.north);
\end{tikzpicture}\Big)^* \ ,
\end{align}
and the mode numbers $r_i$, $s_i$ can be read of from eq.~(\ref{eq:q2_action_physical}), and $\bar{\kappa}^{l,c}_{n,b}$ is given by a similar expression. Since we are now considering physical states, exact momentum conservation can (and must) be imposed, and there is therefore no need for a `weighting factor' $\delta^{(w)}(m;n)$. The full matrix element $\gamma_{n,b}^{m,a}$ of $\tilde{\mathcal{C}}$ is then
\begin{equation}
    \gamma_{n,b}^{m,a} = \sum_{l=0}^{w-2}\: \sum_{c=0}^{w-2-l}\: (\kappa^{l,c}_{m,a})^* \kappa^{l,c}_{n,b} \:+\: \sum_{l=0}^{\lfloor \frac{w-3}{2}\rfloor}\:\sum_{c=l+1 }^{w-2-l}\: (\bar{\kappa}^{l,c}_{m,a})^* \bar{\kappa}^{l,c}_{n,b}\ .
\end{equation}
At large $w$, we would expect to find among the eigenstates of this mixing matrix the long states of the form
\begin{equation}\label{ansatz1}
    \Xi^-_{\frac{n}{w}}\psi^-_{\frac{1}{2}-\frac{n}{w}}\ket{w} = -\tfrac{1}{\sqrt{w+1-n}}\sum_{a=n}^w \xi(\tfrac{n}{w};\tfrac{a}{w}) \, \bar{\psi}^-_{\frac{1}{2}+\frac{n}{w}-\frac{a}{w}}\psi^-_{-\frac{1}{2}+\frac{a}{w}}\psi^-_{\frac{1}{2}-\frac{n}{w}}\ket{w} \ ,
\end{equation}
where the coefficients $\xi(\tfrac{n}{w};\tfrac{a}{w})$ are the same as before, see Figure~\ref{fig:Xi_converge}, and this seems indeed to be the case. There are different ways to give evidence for this, and we shall explain them in turn.

\begin{figure}[h]
\centering
\includegraphics[scale=0.3]{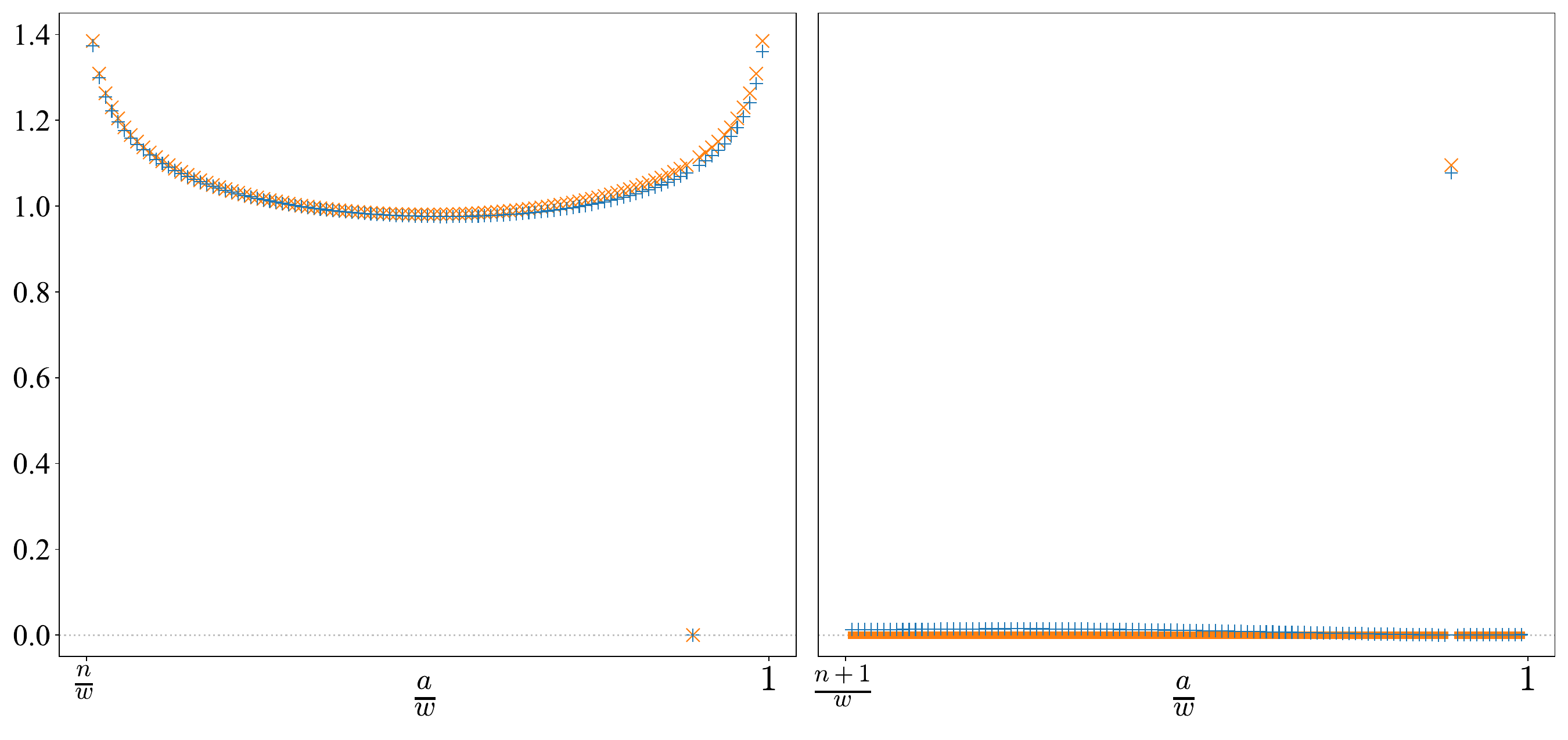}
\caption{A sketch of the identity (\ref{test1}) for $w=120$ and $\tfrac{n}{w}=0.1$: the left-hand-plot shows the left- (blue) and right-hand (orange) side of eq.~(\ref{test1}) for $m=n$; the right-hand-plot depicts both sides for $m= n+1$. The isolated points come from the fact that the basis states vanish for $a=w-m$, and that the states labelled by $(n,a)$ and $(w-a,w-n)$ are identical.}\label{fig:state_density}
\end{figure}

\begin{figure}[h]
\centering
\begin{subfigure}{.49\textwidth}
\includegraphics[scale=0.4]{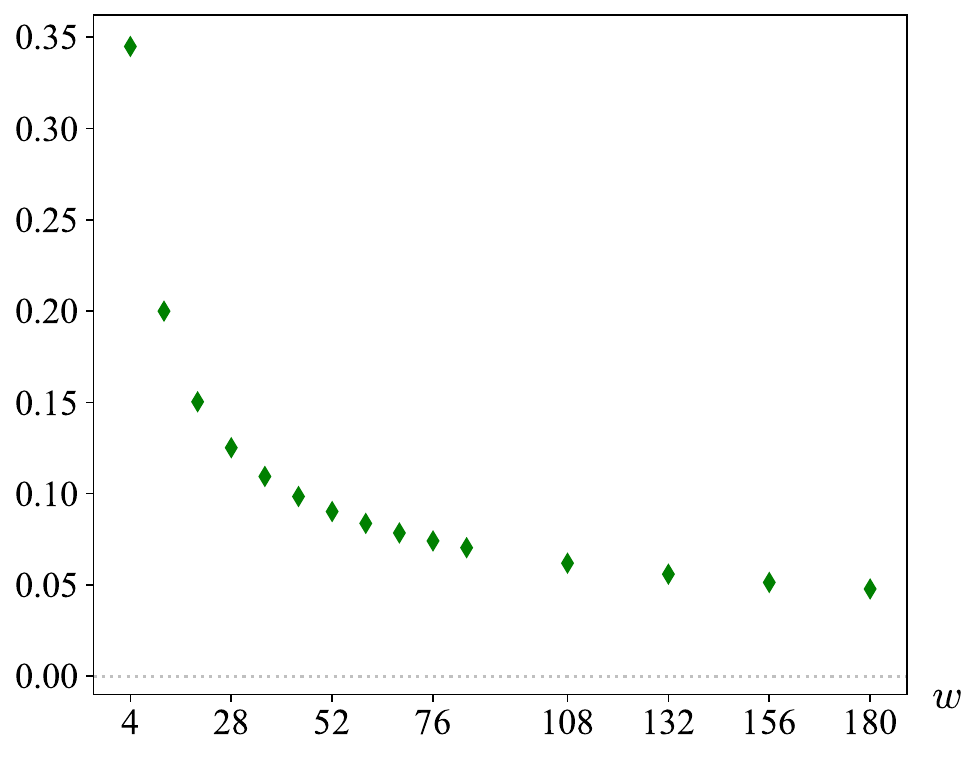}
\end{subfigure}
\begin{subfigure}{.49\textwidth}
\includegraphics[scale=0.4]{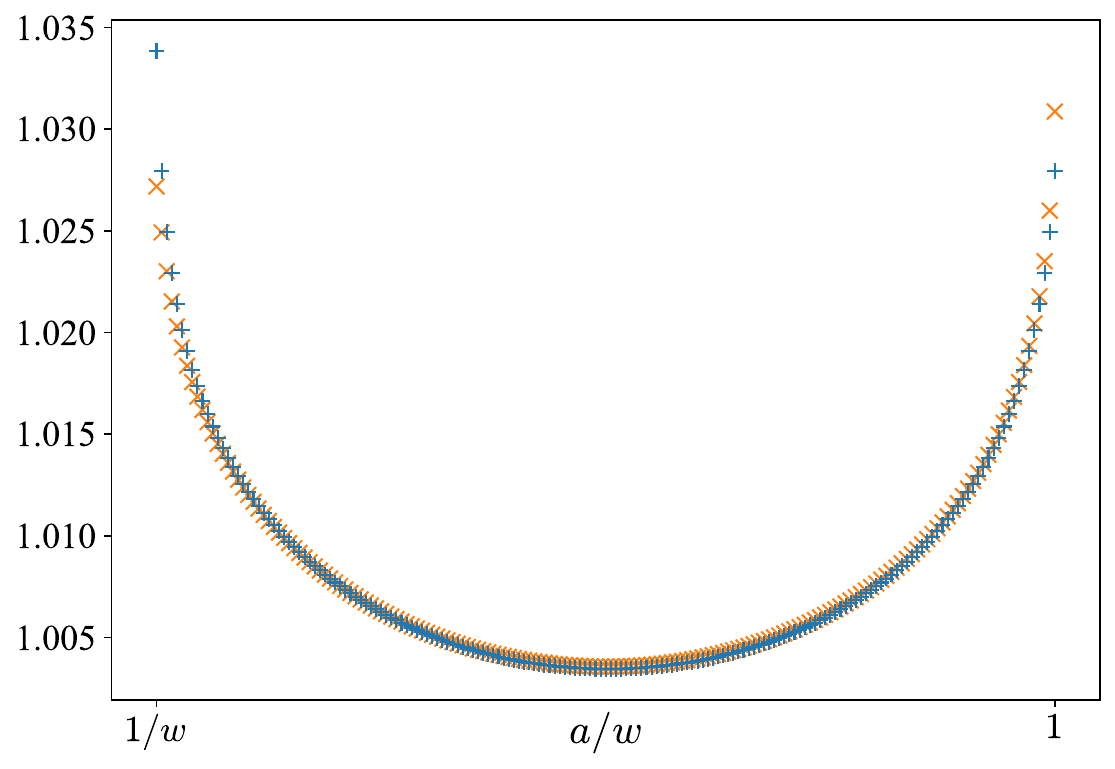}
\end{subfigure}
\caption{On the left: the norm of the state in (\ref{norm1}) for fixed $\tfrac{n}{w}=0.25$ as a function of $w$. On the right: a low energy eigenstate of the negative three-magnon system (orange) compared with the long state $\Xi^-_{\frac{1}{w}}\psi^-_{\frac{1}{2}-\frac{1}{w}}\ket{w}$ (blue)  for $w=180$.}
\label{fig:norm_decay}
\end{figure}

\begin{figure}[h]
\centering
\includegraphics[scale=0.5]{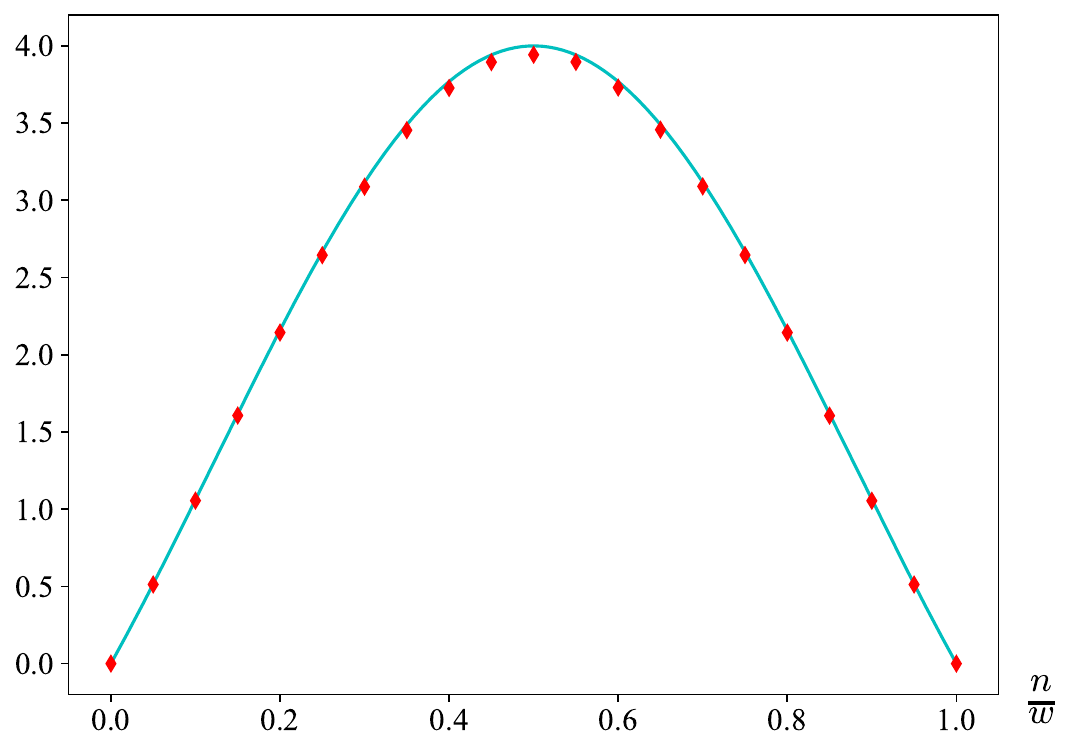}
\caption{The approximate eigenvalues (\ref{eq:approx_ev}) (red diamonds) and the curve of the func\-tion $\epsilon_2(\tfrac{n}{w}) + \epsilon_2(1-\tfrac{n}{w})$ (cyan) for $w=180$ and $\tfrac{n}{w} \in [0,1]$.}
\label{fig:Xi_psi_diag}
\end{figure}

First of all, we can ask whether (\ref{ansatz1}) is at least approximately an eigenvector with the appropriate eigenvalue, i.e.\ we can evaluate
\be\label{CXi}
\tilde{\mcl{C}}\,  \Xi^-_{\frac{n}{w}}\psi^-_{\frac{1}{2}-\frac{n}{w}}\ket{w} = -\sum_{m=0}^{w} \tfrac{1}{\sqrt{w+1-m}}\sum_{a=m}^{w} f_n(\tfrac{m}{w} ;\tfrac{a}{w}) \, \bar\psi^-_{\frac{1}{2} + \frac{m}{w} - \frac{a}{w}} \psi^-_{-\frac{1}{2} + \frac{a}{w}} \, \psi^-_{\frac{1}{2}-\frac{m}{w}}\ket{w} \ ,
\ee
where we have expanded the right-hand-side in terms of a general state of the form (\ref{eq:3mag_neg_basis}). If (\ref{ansatz1}) is an eigenstate we should expect that $f_n(\tfrac{m}{w} ;\tfrac{a}{w})$ is of the form
\be\label{test1}
f_n(\tfrac{m}{w} ;\tfrac{a}{w}) = \delta_{m,n} \bigl( \epsilon_2(\tfrac{n}{w})+\epsilon_2(1-\tfrac{n}{w}) \bigr) \,
\xi(\tfrac{n}{w};\tfrac{a}{w}) \ ,
\ee
and this seems to be approximately true, see Figure~\ref{fig:state_density}.
Another way of exploring the same question is to analyse the $L^2$ norm of the state
\be\label{norm1}
\frac{\tilde{\mathcal{C}}-\bigl(\epsilon_2(\tfrac{n}{w})+\epsilon_2(1-\tfrac{n}{w})\bigr)}{\bigl(\epsilon_2(\tfrac{n}{w})+\epsilon_2(1-\tfrac{n}{w})\bigr)}\:\Xi^-_{\frac{n}{w}}\psi^-_{\frac{1}{2}-\frac{n}{w}}\ket{w} \ ,
\ee
which we expect to go to zero for large $w$; this is plotted in the left panel of Figure~\ref{fig:norm_decay}. We can also compare one of the lowest lightcone energy eigenstates of the three-fermion system (\ref{eq:3mag_neg_basis}) with $\Xi^-_{\frac{1}{w}}\psi^-_{\frac{1}{2}-\frac{1}{w}}\ket{w}$, and this leads to a nice agreement, see the right panel of Figure~\ref{fig:norm_decay}. Finally, we have compared the approximate eigenvalue of $\Xi_{\frac{n}{w}}^- \psi^-_{\frac{1}{2}-\frac{n}{w}}\ket{w}$,
\begin{equation}\label{eq:approx_ev}
\bra{w}\bigl(\Xi_{\frac{n}{w}}^- \psi^-_{\frac{1}{2}-\frac{n}{w}}\bigr)^\dagger \: \tilde{\mcl{C}} \: \Xi_{\frac{n}{w}}^- \psi^-_{\frac{1}{2}-\frac{n}{w}}\ket{w},
\end{equation}
with the curve of $\epsilon_2(\tfrac{n}{w}) + \epsilon_2(1-\tfrac{n}{w})$, see Figure~\ref{fig:Xi_psi_diag}.

\begin{figure}[hb]
\centering
\includegraphics[scale=0.5]{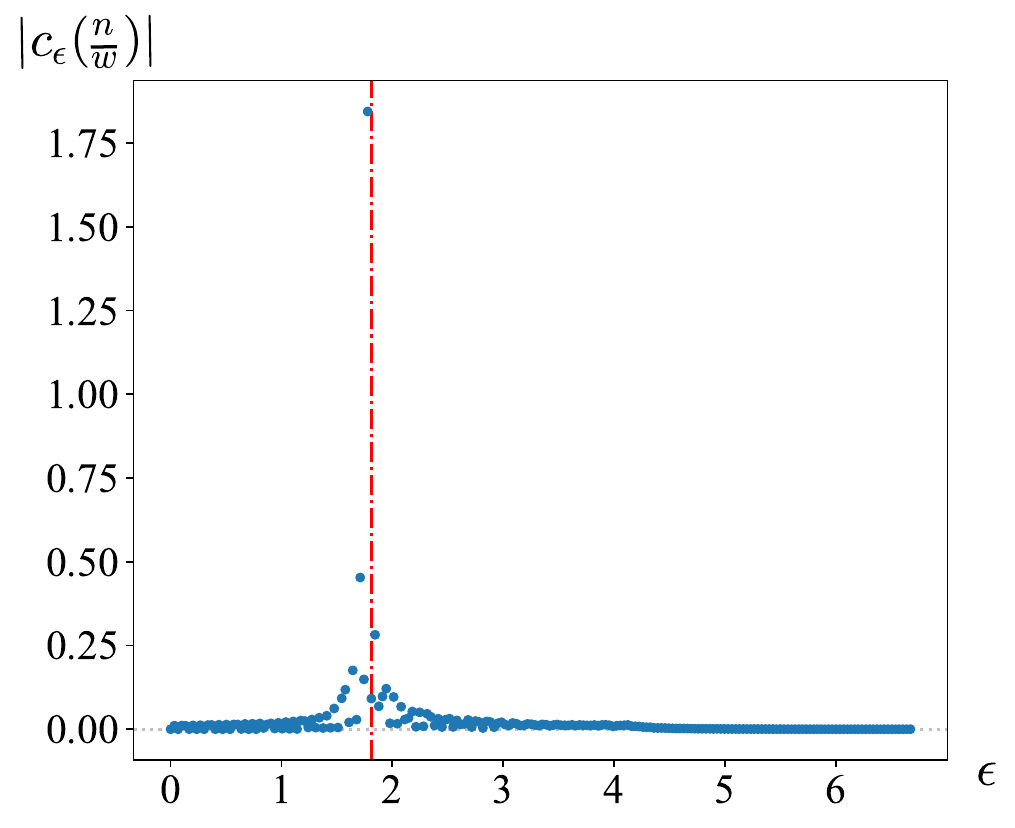}
\caption{Absolute value of the eigenstate basis coefficients $c_{\epsilon}(\tfrac{n}{w})$ of the state $\Xi^-_{\frac{n}{w}}\psi^-_{\frac{1}{2}-\frac{n}{w}}\ket{w}$ for $w=180$ and $n=30$. The red line indicates the expected location $\epsilon = \epsilon_2(\tfrac{n}{w})+\epsilon_2(1-\tfrac{n}{w})$. The range of eigenvalues has been organised in bins of size $\Delta \epsilon \approx 0.03$, and for each bin we have plotted the sum of the absolute values of the coefficients within the bin.}
\label{fig:energy_peak}
\end{figure}

Another useful way of analysing the problem is to expand the states $\Xi^-_{\frac{n}{w}}\psi^-_{\frac{1}{2}-\frac{n}{w}}\ket{w}$, which we expect to be eigenstates at large $w$, in the basis of the exact (numerical) eigenstates $\ket{\epsilon}$ of energy $\epsilon$ at finite $w$,
\begin{equation}
    \Xi^-_{\frac{n}{w}}\psi^-_{\frac{1}{2}-\frac{n}{w}}\ket{w} = \sum_{\epsilon} c_{\epsilon}(\tfrac{n}{w}) \ket{\epsilon}\ .
\end{equation}
The coefficients $c_{\epsilon}(\tfrac{n}{w})$ are sharply peaked at $\epsilon = \epsilon_2(\tfrac{n}{w})+\epsilon_2(1-\tfrac{n}{w})$ as expected, see Figure~\ref{fig:energy_peak}.

\subsection{The full three-fermion spectrum at finite $w$}

On the other hand, analysing the full spectrum of $\tilde{\mathcal{C}}$ on the physical three-magnon states is not so straightforward since there are $\mathcal{O}(w^2)$ states altogether, of which we expect $\mathcal{O}(w)$ `long' eigenstates. Furthermore, the different energy eigenvalues are highly degenerate, and it is therefore difficult to distinguish between `short' and `long' eigenvectors. (For example, two different `long' eigenvectors of approximately the same eigenvalue may just differ by a `short' eigenvector of approximately the same eigenvalue.)

\begin{figure}[h]
\centering

\includegraphics[height=.5\textheight]{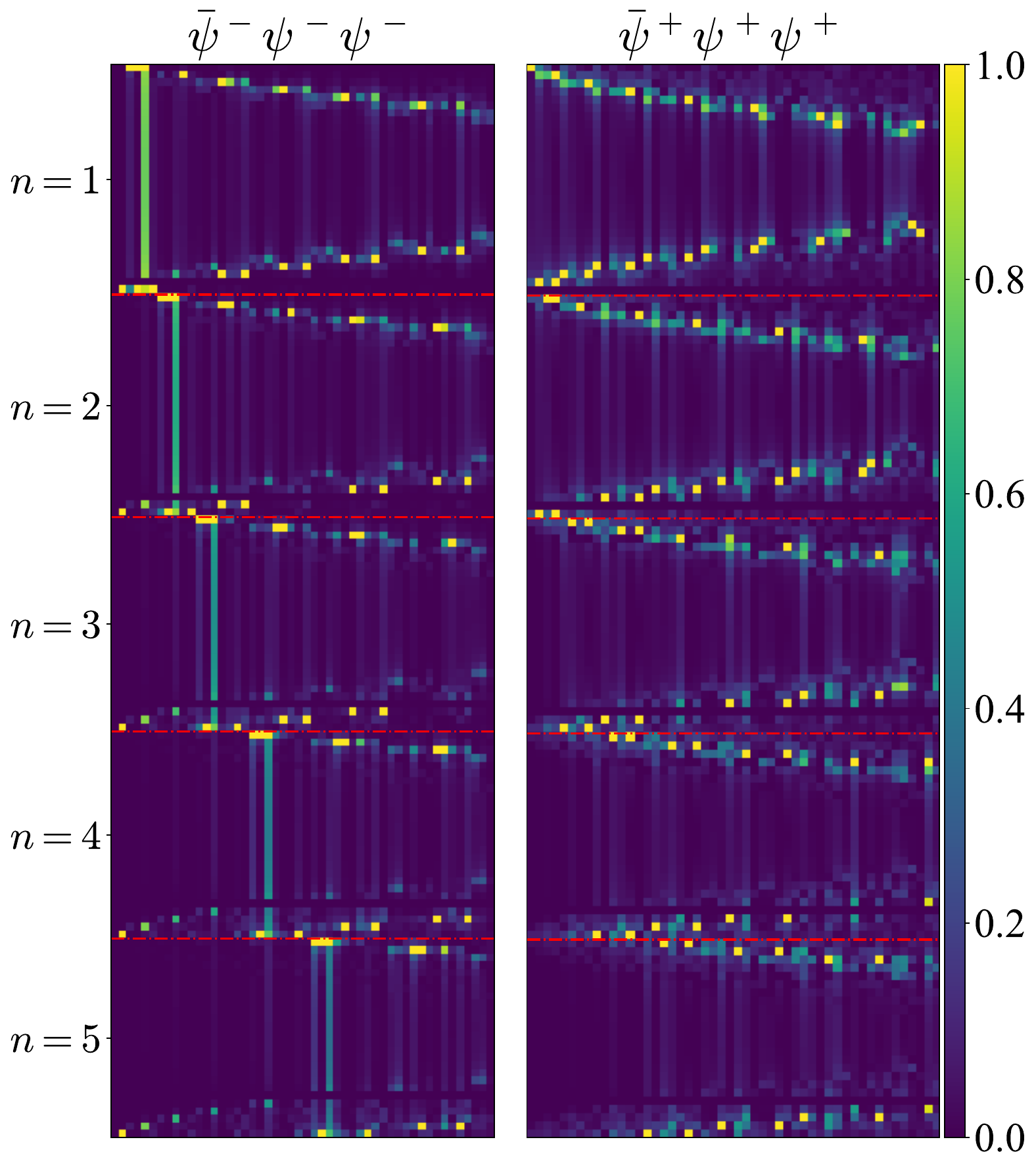}
\caption{The heat-map of the first few low-lying eigenstates in terms of the three-fermion basis states with fixed $n$ for $w=30$. The left-panel shows the situation for the minus sector \eqref{eq:3mag_neg_basis}, whereas on the right the corresponding analysis is done in the plus sector where we consider instead the $\bar{\psi}^+$ and $\psi^+$ fermions, respectively. The solid yellow/green stripes on the left describe `long' eigenstates.}
\label{fig:3mag_minus_spectrum}
\end{figure}

There are a few things, however, we can check. For example, we have drawn the `heat-map', i.e.\ the analogue of Figures~\ref{fig:negative_eigenstates} and \ref{fig:positive_eigenstates}, of the first few low-lying eigenstates of the three-fermion system in terms of the basis states \eqref{eq:3mag_neg_basis}, and we see evidence for `long' strings in the minus sector, but not in the analogous plus-sector (where we replace the fermions by $\bar{\psi}^+$ and $\psi^+$), see Figure~\ref{fig:3mag_minus_spectrum}.

\begin{figure}
\centering
\includegraphics[scale=0.4]{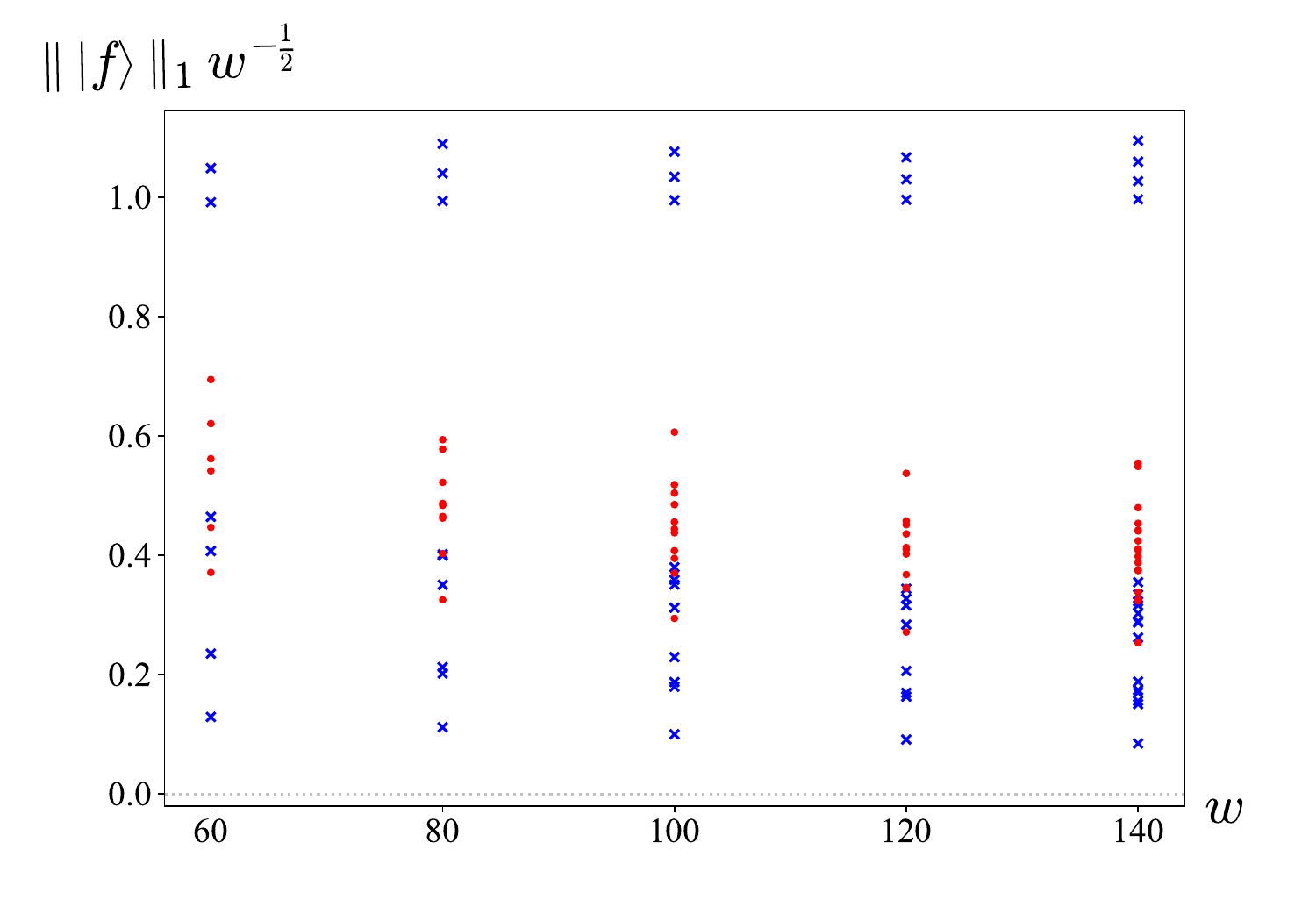}
\caption{The $L^1$ norm \eqref{eq:L1} times the factor $w^{-\frac{1}{2}}$ of the eigenstates with energy $\epsilon \leq 0.25$ of the  negative (blue) and positive (red) three-magnon systems as a function of the cycle length $w$. For $\epsilon \leq 0.25$ there are $2$ long eigenstates for $w=60$, $3$ for $80\leq w \leq 120$ and $4$ for $w=140$ in the minus sector; they are described by the blue dots at the top.}
\label{fig:3mag_L1}
\end{figure}

We have also repeated the $L^1$ norm analysis of eq.~(\ref{L1}) for the fermionic three-magnon sector. Given that there are ${\cal O}(w^2)$ eigenstates, a state that involves all of them more or less uniformly --- this would be a `superlong' state --- would have an $L^1$ norm proportional to $w$, whereas the `long' eigenstates of the form (\ref{ansatz1}) have an $L^1$ norm proportional to $w^{1/2}$, while the `short' states go as $w^0$. Given that there are generically many almost degenerate eigenstates, the distinction between long and short states gets blurred, and thus the analysis is only clean for the first few energy levels (where there are rather few states altogether). Taking $\epsilon \leq 0.25$, the results show again that there are `long' eigenstates in the negative sector, but not in the positive sector, see Figure~\ref{fig:3mag_L1}. Note that, for large $w$, the number of long states in the negative sector agrees with the expected number of states $\Xi^-_{\frac{n}{w}}\psi^-_{\frac{1}{2}-\frac{n}{w}}\ket{w}$ with maximal eigenvalue $0.25$. For example, for $w=140$ there are four long states, and\footnote{Note that the state $\Xi^-_{1-\frac{n}{w}}\psi^-_{-\frac{1}{2}+\frac{n}{w}}\ket{w}$, which has the same eigenvalue as $\Xi^-_{\frac{n}{w}}\psi^-_{\frac{1}{2}-\frac{n}{w}}\ket{w}$, is not a true long state when $n$ is small because it involves only $n+1$ modes.}
\begin{equation}
   \epsilon_2(\tfrac{n}{140}) + \epsilon_2(1-\tfrac{n}{140}) < 0.25,\quad n=0,\:1,\:2,\:3 \ , \qquad
    \epsilon_2(\tfrac{4}{140}) + \epsilon_2(1-\tfrac{4}{140}) \approx 0.29 \ .
\end{equation}
Additionally, we have not found any `superlong' eigenstates. We have also checked that all long eigenstates with momentum $0\leq \tfrac{n}{w}\leq 0.1$ are well approximated by the corresponding state $\Xi^-_{\frac{n}{w}}\psi^-_{\frac{1}{2}-\frac{n}{w}}\ket{w}$;\footnote{Long states with $\tfrac{n}{w}\leq \tfrac{1}{2}$ have a discontinuity at $a=w-n$, where the basis state vanishes. Near the discontinuity, the coefficients deviate from the expected curve $\xi(\tfrac{n}{w};\tfrac{a}{w})$, but this difference tends to zero for $w\to\infty$.}  in particular, this demonstrates that the only long eigenstates involve $\Xi^-$.

\subsection{Higher magnon checks}

\begin{figure}[ht]
\centering
\includegraphics[scale=0.4]{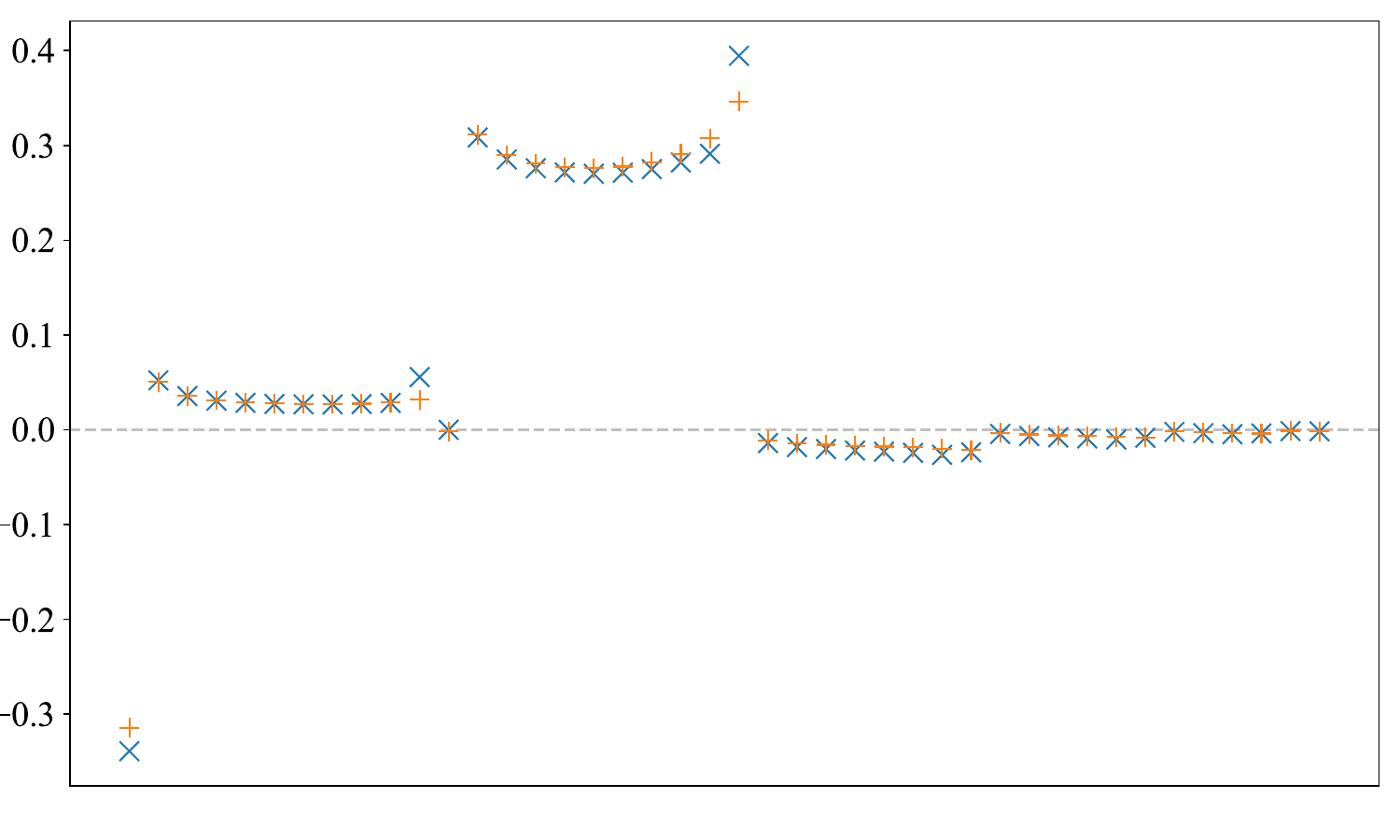}
\caption{Coefficients of the three-magnon part of a long eigenstate at $w=12$. The calculation including only three-magnon states (blue crosses) agrees well with the calculation with added five-magnon states (orange pluses).}\label{fig:physical_comparison}
\end{figure}

As for the checks on the unphysical calculation carried out in Section \ref{sec:unphysical_higher_magnon}, we can again compare these results to a more accurate calculation including five-magnon states, using the techniques of \cite{Gaberdiel:2024nge}. We have done this calculation for $w=12$, where instead of the $42$ states that mix when we only consider three-magnon states, we now have to deal with $1778$ states. We then project onto the three-magnon part of the eigenstates, i.e.\ we pick the eigenstates with the largest overlap with the three-magnon states, and we observe again that these states are either short or long in the sense discussed in Section~\ref{sec:minus}. In Figure~\ref{fig:physical_comparison}, we compare a low lying long eigenstate found in this analysis with one obtained by only including three-magnon states above. One can see that they agree very well.

\section{Emergent directions}\label{sec:emergent_directions}

In the previous sections we have found strong evidence for the existence of collective modes that give rise to `long' eigenstates. In this section we want to argue that they correspond to the $\text{AdS}_3\times {\rm S}^3$ excitations in the dual AdS string theory.
In fact, it was already suggested in \cite{Lunin:2002fw,Gomis:2002qi,Gava:2002xb} that the BMN excitations \cite{Berenstein:2002jq} associated to the ${\rm AdS}_3$ and ${\rm S}^3$ directions should be identified with the fractional ${\cal N}=4$ modes in the symmetric orbifold. Recall that in the pp-wave limit of $\text{AdS}_3\times {\rm S}^3\times \mbb{T}^4$, one $\text{AdS}_3$ and one ${\rm S}^3$ direction are chosen as the  light-cone directions. The string excitations along the remaining directions on $\text{AdS}_3$ and ${\rm S}^3$ can then be parametrised by the oscillators
\begin{equation}
a^{\text{AdS}}(n)\ ,\quad b^{\text{AdS}}(n)\ ,\quad a^{\rm S}(n)\ ,\quad b^{\rm S}(n)\ ,\qquad n\in \mbb{Z} \ ,
\end{equation}
and it follows from the BMN analysis  \cite{Berenstein:2002jq} that their dispersion relations are of the form \cite{Berenstein:2002jq,Hoare:2013pma}\footnote{In \cite{Berenstein:2002jq} only the $+$ solution was given; it was subsequently realised in \cite{Hoare:2013pma} that this is only true for half the modes.}
\begin{equation}\label{eq:BMN_dispersion}
\sqrt{\sin^2\alpha + \left(\cos\alpha \pm \frac{n}{\alpha' p^+}\right)^2}\ ,
\end{equation}
where $\cos\alpha$ parametrises the ratio of NS-NS and R-R fluxes,
\begin{equation}
\cos\alpha = \frac{q_{\text{NS}}}{\sqrt{q^2_{\text{NS}}+g_s^2 q^2_{\text{R}}}}\ ,
\end{equation}
and for the $a$ ($b$) oscillators we pick the minus (plus) sign in the dispersion relation eq.~(\ref{eq:BMN_dispersion}). (Note that the mode number $n$ can also be negative.)

It was suggested in \cite{Lunin:2002fw,Gomis:2002qi,Gava:2002xb} that these BMN modes should be identified with the fractional ${\cal N}=4$ modes as,\footnote{Since the dictionary of \cite{Lunin:2002fw,Gomis:2002qi,Gava:2002xb} was based on the old relations of \cite{Berenstein:2002jq}, it is actually not quite consistent; the correct version was worked out in \cite{Gaberdiel:2023lco}.}
\begin{align}\label{oldiden}
a^{\text{AdS}}(n) &\longleftrightarrow L_{-1+\frac{n}{w}}\ , & b^{\text{AdS}}(n) &\longleftrightarrow \tilde{L}_{-1-\frac{n}{w}}\ ,\nonumber\\
a^{S}(n) &\longleftrightarrow K^-_{\frac{n}{w}}\ , & b^{S}(n) &\longleftrightarrow \tilde{K}^-_{-\frac{n}{w}}\ ,
\end{align}
where the tilde refers to the right-movers; the analysis for the fermionic generators is similar: they correspond to the fractional $G^-$ and $G'^{-}$ modes, resp.\ their right-moving analogues. The main evidence for this was that the zero modes of these generators reproduce the appropriate global symmetries of the background, but at the time it was not possible to reproduce the dispersion relation (\ref{eq:BMN_dispersion}) from the symmetric orbifold. It was also not clear whether these modes are really independent from the individual torus modes that correspond to the torus excitations in the spacetime. (After all, the ${\cal N}=4$ modes can be expressed in terms of the bilinears of the torus modes.)

It should now be clear how our results fit into this picture: if we make the identifications, see e.g.\ \cite[Section 6.2]{Gaberdiel:2023lco},
\begin{equation}
2\pi g = g_s q_{\text{R}}\ ,\qquad p(n) = \frac{1}{1+4\pi^2 g^2}\frac{n}{\alpha' p^+}\ ,
\end{equation}
and expand the dispersion relation in $g$ and for $\lvert p(n) \rvert \ll 1$ we reproduce precisely the BMN dispersion relation (\ref{eq:BMN_dispersion}) to order ${\cal O}(p^2)$, provided we make the identifications\footnote{This calculation was already performed in  \cite{Gaberdiel:2023lco}; at the time, we believed that the original proposal of (\ref{oldiden}) was correct, and that the fractional ${\cal N}=4$ modes also satisfy the dispersion relation (\ref{dispersion}).}
\begin{align}
a^{\text{AdS}}(n) &\longleftrightarrow \Lambda_{-1+p(n)}\ , & b^{\text{AdS}}(n) &\longleftrightarrow \tilde{\Lambda}_{-1-p(n)}\ ,\nonumber\\
a^{\rm S}(n) &\longleftrightarrow \Xi^-_{p(n)}\ , & b^{\rm S}(n) &\longleftrightarrow \tilde{\Xi}^-_{-p(n)}\ .
\end{align}
We therefore claim that our `long' magnon modes describe precisely the $\text{AdS}_3\times {\rm S}^3$ excitations!\footnote{The `short' magnon modes correspond, on the other hand, to the spacetime $\mathbb{T}^4$ modes \cite{Gaberdiel:2023lco,Frolov:2023pjw}.}

A few comments are in order. While, our proposal differs from (\ref{oldiden}), the symmetry argument still applies since our long magnons reduce to the ${\cal N}=4$ generators for integer mode number. (However, for non-integer $\frac{n}{w}$, the two expressions differ.) Furthermore, given that the perturbation lifts the degeneracy of states, it is clear that these modes are indeed independent from the torus modes, thus vindicating the intuition of  \cite{Lunin:2002fw,Gomis:2002qi,Gava:2002xb}. This independence only emerges in the large $w$ limit, but this is also the limit to which the BMN analysis applies. Finally, it is very reassuring that our analysis \emph{only} finds these long magnon modes, but not for example a long magnon mode in the positive sector that includes the fractional $K^+$ modes --- there are no additional modes in the BMN analysis  to which they could have corresponded to! Finally, by construction it is clear that the `long' modes are also present at zero coupling, i.e.\ even the free symmetric orbifold contains them naturally. (However, it is difficult to single them out among all the degenerate eigenstates at the free point.)

\section{Discussion and Conclusion}\label{sec:concl}

In this paper we have given very strong evidence for the assertion that the ${\rm AdS}_3 \times {\rm S}^3$ magnons of the BMN limit are already present in the dual symmetric orbifold. Our proposal follows in spirit the old idea of \cite{Lunin:2002fw,Gomis:2002qi,Gava:2002xb}, but the details are a bit different in that the relevant symmetric orbifold modes are not just the fractional ${\cal N}=4$ modes, but are instead given by, e.g.\ eq.~(\ref{eq:Xi_op}). Furthermore, relative to \cite{Lunin:2002fw,Gomis:2002qi,Gava:2002xb} we have also managed to reproduce the BMN dispersion relation directly from the symmetric orbifold.

Our analysis relied heavily on the recent progress with our understanding of the perturbation of the symmetric orbifold \cite{Gaberdiel:2023lco}, see also \cite{Gaberdiel:2024nge}. However, even using the simplifying assumptions of \cite{Gaberdiel:2023lco} --- including the additional contributions of \cite{Gaberdiel:2024nge} complicates the analysis obviously even further --- the calculation is formidable, and we have only managed to explore various simple cases numerically. It would obviously be very interesting to find a more analytical approach to this problem. This should also help to identify the specific form of the fractional `long' magnon modes, i.e.\ the functional form of the coefficients $\xi(\tfrac{n}{w};\tfrac{a}{w})$ in, say eq.~(\ref{eq:Xi_op}), see Figure~\ref{fig:Xi_converge}. Furthermore, it would be reassuring to confirm that these `long' modes also behave as expected from integrablity, i.e.\ that the action of the right-moving supercharges on the `long' left modes has a similar form as eq.~(\ref{commu}) to leading order in $1/w$.

\section*{Acknowledgements} This paper is partially based on the Master thesis of one of us (D.K.). We thank Ofer Aharony, Simon Ekhammar, Rajesh Gopakumar, Bin Guo, Anthony Houppe, Oleg Lunin, Kiarash Naderi, and Vit Sriprachyakul for useful conversations.  The work of BN is supported through a personal grant of MRG from the Swiss National Science Foundation. The work of the group at ETH is also supported in part by the Simons Foundation grant 994306 (Simons Collaboration on Confinement and QCD Strings), as well as the NCCR SwissMAP that is also funded by the Swiss National Science Foundation.

\appendix

\section{Conventions}\label{app:conventions}

\begingroup
\allowdisplaybreaks

In this appendix we collect our conventions for the torus fields and the ${\cal N}=4$ superconformal generators.

\subsection{\texorpdfstring{$\mcl{N}=4$ algebra}{N=4 algebra}}\label{app:algebra}

We denote the left-moving four bosons and four fermions of the $\mbb{T}^4$ by $\alpha^i,\bar{\alpha}^i$, $i=1,2$ and $\psi^\pm,\bar{\psi}^\pm$, respectively. They satisfy the OPE relations
\begin{equation}
\bar{\alpha}^i(x)\alpha^j(y)\sim \frac{\epsilon^{ij}}{\big(x-y\big)^2}\ ,\qquad \bar{\psi}^\pm(x)\psi^\mp(y)\sim \frac{\pm 1}{x-y} \ .
\end{equation}
Right-movers are always denoted by a tilde. The (left-moving) $\mcl{N}=4$ generators are built out of these fields as
\begin{equation}\label{N4fields}
\begin{array}{cclccl}
G^+ &= & \bar{\alpha}^2\,\psi^++\alpha^2\,\bar{\psi}^+\ , \qquad & K^+ &= & \bar{\psi}^+\,\psi^+\ , \\
G'^+ &= & -\bar{\alpha}^1\,\psi^+-\alpha^1\,\bar{\psi}^+\ ,\qquad & K^- &= & -\bar{\psi}^-\,\psi^-\ , \\
G^- &= & \bar{\alpha}^1\,\psi^-+\alpha^1\,\bar{\psi}^-\ , \qquad & K^3 &= & \frac{1}{2}:\bar{\psi}^+\,\psi^-+\bar{\psi}^-\,\psi^+:\ , \\
G'^- &= & \bar{\alpha}^2\,\psi^-+\alpha^2\,\bar{\psi}^-\ , \qquad & &&
\end{array}
\end{equation}
and
\begin{equation}
L =  \,:\bar{\alpha}^1\,\alpha^2-\bar{\alpha}^2\,\alpha^1:+\,\frac{1}{2}:\psi^+\,\partial \bar{\psi}^- + \bar{\psi}^-\,\partial \psi^+ - \bar{\psi}^+\,\partial\psi^- - \psi^-\,\partial\bar{\psi}^+:\ .
\end{equation}
These fields generate the $c=6$ ${\cal N}=4$ superconformal algebra, whose modes satisfy
\begin{equation}
\begin{array}{rcl}
{}   [L_m,L_n]&=& (m-n)L_{m+n}+\tfrac{1}{2}(m^3-m)\delta_{m+n,0}\ ,\\
{}   [L_m,K^a_n]&= & -nK^a_{m+n}\ ,\\
{}   [L_m,G^a_r]&= & \big(\tfrac{m}{2}-r\big)G^a_{m+r}\ ,\\
{}   [L_m,G'^a_r]&= & \big(\tfrac{m}{2}-r\big)G'^a_{m+r}\ ,\\
{}   \{G^+_r,G^-_s\}&= & \{G'^+_r,G'^-_s\}=L_{r+s}+(r-s)K^3_{r+s}+\big(r^2-\tfrac{1}{4}\big)\delta_{r+s,0}\ ,\\
{}   \{G^\pm_r,G'^\pm_s\}&= & \mp(r-s)K^\pm_{r+s}\ ,
\end{array}
\end{equation}
as well as
\begin{equation}
\begin{array}{cclccl}
{}[K^3_m,K^3_n] &=& \tfrac{m}{2}\delta_{m+n,0}\ , \qquad   & [K^3_m,G^\pm_r]&=& \pm\tfrac{1}{2}G^\pm_{m+r}\ , \\
{} [K^3_m,K^\pm_n]&=& \pm K^\pm_{m+n}\ ,\qquad & [K^3_m,G'^\pm_r]&= & \pm\tfrac{1}{2}G'^\pm_{m+r}\ , \\
{}[K^+_m,K^-_n]&= & 2K^3_{m+n}+m\,\delta_{m+n,0}\ ,\qquad & & &
\end{array}
\end{equation}
and
\begin{equation}
\begin{array}{cclccl}
{}[K^+_m,G^-_r] &=& G'^+_{m+r}\ , \qquad & [K^+_m,G'^-_r] &=& -G^+_{m+r}\ ,\\
{}[K^-_m,G^+_r] &=& -G'^-_{m+r}\ , \qquad & [K^-_m,G'^+_r] &=& G^-_{m+r}\ .
\end{array}
\end{equation}
The other (anti-)commutators vanish. It is sometimes convenient to bosonise the free fermions as
\begin{equation}
\psi^+=e^{i\phi}\ ,\quad \bar{\psi}^- =-e^{-i\phi}\ ,\quad \bar{\psi}^+ = e^{i\phi'}\ ,\quad \psi^- = e^{-i\phi'}\ ,
\end{equation}
where $\phi,\phi'$ are free bosons with OPEs
\begin{equation}
\phi(x)\phi(y)\sim -\log(x-y)\ ,\quad \phi'(x)\phi'(y)\sim -\log(x-y)\ .
\end{equation}

\endgroup

\subsection{States}\label{app:states}

In the symmetric product orbifold of $\mbb{T}^4$, we denote the $w$-twisted sector ground-state by $\sigma_w$. For odd $w$ it has conformal dimension $h=\tilde{h}=\tfrac{w^2-1}{4w}$ and $\mfr{su}(2)$ charge $m=\tilde{m} = 0$. For even $w$, there is actually a doublet of Ramond ground states, and we denote by $\sigma_w$ the state with $h=\tilde{h}=\tfrac{w}{4}$ and $\mfr{su}(2)$ charge  $m=\tilde{m}=-\tfrac{1}{2}$.

In each twisted sector, there are four (left and right) BPS states with charge and dimension $\frac{w-1}{2},\frac{w}{2},\frac{w+1}{2}$. We denote the BPS state with $h=\tilde{h}=m=\tilde{m}=\tfrac{w-1}{2}$ by $\ket{\text{BPS}_-}_w$. Explicitly, this state is
\begin{align}
\ket{\mrm{BPS}_-}_w &= \big(\bar{\psi}^+_{-\frac{w-2}{2w}}\psi^+_{-\frac{w-2}{2w}}\cdots\bar{\psi}^+_{-\frac{1}{2w}}\psi^+_{-\frac{1}{2w}}\big)\big(\tilde{\bar{\psi}}^+_{-\frac{w-2}{2w}}\tilde{\psi}^+_{-\frac{w-2}{2w}}\cdots\tilde{\bar{\psi}}^+_{-\frac{1}{2w}}\tilde{\psi}^+_{-\frac{1}{2w}}\big)\sigma_w\ ,\quad\text{for $w$ odd}\ ,\nonumber \\
\ket{\mrm{BPS}_-}_w &= \big(\bar{\psi}^+_{-\frac{w-2}{2w}}\psi^+_{-\frac{w-2}{2w}}\cdots\bar{\psi}^+_{0}\psi^+_{0}\big)\big(\tilde{\bar{\psi}}^+_{-\frac{w-2}{2w}}\tilde{\psi}^+_{-\frac{w-2}{2w}}\cdots\tilde{\bar{\psi}}^+_{0}\tilde{\psi}^+_{0}\big)\sigma_w\ ,\quad\text{for $w$ even}\ . \label{bottomBPS}
\end{align}
The other BPS states can be obtained by applications of $\bar{\psi}^+_{-1/2}$, $\psi^+_{-1/2}$ and the corresponding right-movers. The standard BPS state we work with has $h=\tilde{h}=m=\tilde{m}=\frac{w+1}{2}$ and is given by
\begin{equation}
\ket{w} ={\psi}^+_{-\frac{1}{2}}\bar{\psi}^+_{-\frac{1}{2}}\tilde{{\psi}}{}^+_{-\frac{1}{2}}\tilde{\bar{\psi}}^+_{-\frac{1}{2}}\ket{\mrm{BPS}_-}_w\ .
\end{equation}
On the covering surface, this state corresponds to
\begin{equation}
    :e^{i\tfrac{w+1}{2}(\phi+\phi'+\tilde{\phi}+\tilde{\phi}')}:\ ,
\end{equation}
in the bosonised form of the free fermions.

The perturbing field $\Phi$ is a descendant of the lower BPS state in the two-twisted sector, given by
\begin{equation}\label{perturbing}
    \ket{\Phi} = \frac{i}{\sqrt{2}}\,\big(G^-_{-\frac{1}{2}}\tilde{G}'^-_{-\frac{1}{2}}-G'^-_{-\frac{1}{2}}\tilde{G}^-_{-\frac{1}{2}}\big)\ket{\mrm{BPS}_-}_2\ .
\end{equation}
This field has $h=\tilde{h}=1$ and is a singlet with respect to the left- and right-moving $\mfr{su}(2)$.

\subsection{Contractions}\label{app:contractions}

Here, we give explicit expressions for the contractions used in calculating the anomalous dimension matrices. These are transitions from the $w$- to the $w+1$-twisted sector. Transitions to the $w-1$-twisted sector can be obtained by conjugation.

Firstly, we have the  ``crossed'' contractions,\footnote{We express the contractions in terms of binomial coefficients instead of Gamma functions, as was done in \cite{Gaberdiel:2023lco}, as this is more suitable for numerical evaluation. We also use a different notation for the mode numbers of the magnons.} where a fermion can contract with a boson, mediated by the perturbation. We denote this by a crossed bracket below the magnons:

\begin{align}
\begin{tikzpicture}[baseline=-.5ex]
    \node[](n1)at(0,0){$\bar{\psi}^+_{\frac{1}{2}-\frac{m}{w+1}}$};
    \node[](n2)[right=-1ex of n1]{$\alpha^2_{-1+\frac{n}{w}}$};
    \draw[](n1.south)-- ([yshift=-1ex]n1.south)--([yshift=-1.25ex]n2.south)node[midway]{$\times$}--(n2.south);
\end{tikzpicture} &= e^{-i\pi \frac{m}{w+1}}(w+1)^{\frac{n}{w}}w^{1-\frac{m}{w+1}}\left(-\tfrac{w}{w+1}\right)^{m-n-1}\sqrt{1-\tfrac{n}{w}} \nonumber\\
&\qquad\times \binom{-1+\frac{n}{w}}{w-n}\,\binom{-\frac{m}{w+1}}{w+1-m}\ ,\\
&\nonumber\\
\begin{tikzpicture}[baseline=-.5ex]
    \node[](n1)at(0,0){$\bar{\alpha}^2_{1-\frac{m}{w+1}}$};
    \node[](n2)[right=-1ex of n1]{$\psi^+_{-\frac{3}{2}+\frac{n}{w}}$};
    \draw[](n1.south)-- ([yshift=-1.1ex]n1.south)--([yshift=-1ex]n2.south)node[midway]{$\times$}--(n2.south);
\end{tikzpicture} &= \sqrt{\frac{1-\frac{m}{w+1}}{1-\frac{n}{w}}} \begin{tikzpicture}[baseline=-.5ex]
    \node[](n1)at(0,0){$\bar{\psi}^+_{\frac{1}{2}-\frac{m}{w+1}}$};
    \node[](n2)[right=-1ex of n1]{$\alpha^2_{-1+\frac{n}{w}}$};
    \draw[](n1.south)-- ([yshift=-1ex]n1.south)--([yshift=-1.25ex]n2.south)node[midway]{$\times$}--(n2.south);
\end{tikzpicture}.
\end{align}

The other crossed contractions required for the calculation can be expressed in terms of these as
\begin{equation}
\begin{tikzpicture}[baseline=-.5ex]
    \node[](n1)at(0,0){$\psi^+_{q}$};
    \node[](n2)[right=-1ex of n1]{$\bar{\alpha}^2_{p}$};
    \draw[](n1.south)-- ([yshift=-1ex]n1.south)--([yshift=-1ex]n2.south)node[midway]{$\times$}--(n2.south);
\end{tikzpicture} = - \begin{tikzpicture}[baseline=-.5ex]
    \node[](n1)at(0,0){$\bar{\psi}^+_{q}$};
    \node[](n2)[right=-1ex of n1]{$\alpha^2_{p}$};
    \draw[](n1.south)-- ([yshift=-1ex]n1.south)--([yshift=-1ex]n2.south)node[midway]{$\times$}--(n2.south);
\end{tikzpicture}\ ,\quad
\begin{tikzpicture}[baseline=-.5ex]
    \node[](n1)at(0,0){$\alpha^2_{q}$};
    \node[](n2)[right=-1ex of n1]{$\bar{\psi}^+_{p}$};
    \draw[](n1.south)-- ([yshift=-1.1ex]n1.south)--([yshift=-1.1ex]n2.south)node[midway]{$\times$}--(n2.south);
\end{tikzpicture} = - \begin{tikzpicture}[baseline=-.5ex]
    \node[](n1)at(0,0){$\bar{\alpha}^2_{q}$};
    \node[](n2)[right=-1ex of n1]{$\psi^+_{p}$};
    \draw[](n1.south)-- ([yshift=-1.1ex]n1.south)--([yshift=-1.1ex]n2.south)node[midway]{$\times$}--(n2.south);
\end{tikzpicture}\ .
\end{equation}

The second type of contractions appearing in the calculation are ``same-species'' contractions, denoted by a bracket above the magnons. The expressions are

\begin{align}
\begin{tikzpicture}[baseline=-.5ex]
    \node[](n1)at(0,0){$\bar{\psi}^+_{\frac{1}{2}-\frac{m}{w+1}}$};
    \node[](n2)[right=-1ex of n1]{$\psi^-_{-\frac{1}{2}+\frac{n}{w}}$};
    \draw[](n1.north)-- ([yshift=1ex]n1.north)--([yshift=1ex]n2.north)--(n2.north);
\end{tikzpicture}
    &=-e^{-i\pi \frac{m}{w+1}}(w+1)^{-\frac{1}{2}+\frac{n}{w}}w^{\frac{1}{2}-\frac{m}{w+1}}\left(-\tfrac{w}{w+1}\right)^{m-n-1}\left(1-\tfrac{n}{w}\right)\nonumber\\
    & \qquad \times\frac{1}{\frac{m}{w+1}- \frac{n}{w}}\, \binom{-1+\frac{n}{w}}{w-n}\,\binom{-\frac{m}{w+1}}{w+1-m}\ ,
\end{align}
for general momenta $\tfrac{m}{w+1}\neq \tfrac{n}{w}$, and
\begin{align}
    \begin{tikzpicture}[baseline=-.5ex]
    \node[](n1)at(0,0){$\bar{\psi}^+_{\frac{1}{2}-k}$};
    \node[](n2)[right=-1ex of n1]{$\psi^-_{-\frac{1}{2}+k}$};
    \draw[](n1.north)-- ([yshift=1ex]n1.north)--([yshift=1ex]n2.north)--(n2.north);
\end{tikzpicture} = \begin{cases}
    \sqrt{\tfrac{w+1}{w}}, \quad k=1\ , \\
    \sqrt{\tfrac{w}{w+1}}, \quad k\leq 0\ ,
\end{cases}
\end{align}
when the momentum is an integer $k=\tfrac{m}{w+1}=\tfrac{n}{w}$. Furthermore,
\begin{align}
    \begin{tikzpicture}[baseline=-.5ex]
    \node[](n1)at(0,0){$\bar{\psi}^-_{\frac{3}{2}-\frac{m}{w+1}}$};
    \node[](n2)[right=-1ex of n1]{$\psi^+_{-\frac{3}{2}+\frac{n}{w}}$};
    \draw[](n1.north)-- ([yshift=1ex]n1.north)--([yshift=1ex]n2.north)--(n2.north);
\end{tikzpicture}
    &= \frac{1-\frac{m}{w+1}}{1-\frac{n}{w}} \begin{tikzpicture}[baseline=-.5ex]
    \node[](n1)at(0,0){$\bar{\psi}^+_{\frac{1}{2}-\frac{m}{w+1}}$};
    \node[](n2)[right=-1ex of n1]{$\psi^-_{-\frac{1}{2}+\frac{n}{w}}$};
    \draw[](n1.north)-- ([yshift=1ex]n1.north)--([yshift=1ex]n2.north)--(n2.north);
\end{tikzpicture},
\end{align}
as well as
\begin{equation}
\begin{tikzpicture}[baseline=-.5ex]
    \node[](n1)at(0,0){$\psi^+_{q}$};
    \node[](n2)[right=-1ex of n1]{$\bar{\psi}^-_{p}$};
    \draw[](n1.north)-- ([yshift=1ex]n1.north)--([yshift=1ex]n2.north)--(n2.north);
\end{tikzpicture} =
\begin{tikzpicture}[baseline=-.5ex]
    \node[](n1)at(0,0){$\bar{\psi}^+_{q}$};
    \node[](n2)[right=-1ex of n1]{$\psi^-_{p}$};
    \draw[](n1.north)-- ([yshift=1ex]n1.north)--([yshift=1ex]n2.north)--(n2.north);
\end{tikzpicture} \ ,\quad
\begin{tikzpicture}[baseline=-.5ex]
    \node[](n1)at(0,0){$\psi^-_{q}$};
    \node[](n2)[right=-1ex of n1]{$\bar{\psi}^+_{p}$};
    \draw[](n1.north)-- ([yshift=1ex]n1.north)--([yshift=1ex]n2.north)--(n2.north);
\end{tikzpicture} =
\begin{tikzpicture}[baseline=-.5ex]
    \node[](n1)at(0,0){$\bar{\psi}^-_{q}$};
    \node[](n2)[right=-1ex of n1]{$\psi^+_{p}$};
    \draw[](n1.north)-- ([yshift=1ex]n1.north)--([yshift=1ex]n2.north)--(n2.north);
\end{tikzpicture}\ .
\end{equation}
The weighting function $\delta^{(w)}(m;n)$ in eq.~(\ref{eq:q2_action}) is a boson same-species contraction,
\be\label{deltamn}
    \delta^{(w)}(m;n) = \Big(
\begin{tikzpicture}[baseline=-.5ex]
    \node[](n1)at(0,0){$\bar{\alpha}^1_{1+\frac{n}{w}}$};
    \node[](n2)[right=-1ex of n1]{$\alpha^2_{-1-\frac{m}{w-1}}$};
    \draw[](n1.north)-- ([yshift=1ex]n1.north)--([yshift=1ex]n2.north)--(n2.north);
\end{tikzpicture}\Big)^*= \sqrt{\frac{1+\frac{n}{w}}{1+\frac{m}{w-1}}} \Big(\begin{tikzpicture}[baseline=-.5ex]
    \node[](n1)at(0,0){$\bar{\psi}^+_{\frac{1}{2}+\frac{n}{w}}$};
    \node[](n2)[right=-1ex of n1]{$\psi^-_{-\frac{1}{2}-\frac{m}{w-1}}$};
    \draw[](n1.north)-- ([yshift=1ex]n1.north)--([yshift=1ex]n2.north)--(n2.north);
\end{tikzpicture}\Big)^*,
\ee
which mimics the presence of an additional ``inert'' boson $\alpha^2_{-1-\frac{n}{w}}$ in the two-fermion basis states (\ref{eq:minus_space}) that only contracts with another boson $\alpha^2_{-1-\frac{m}{w-1}}$ in the $w-1$-twisted cycle sector (\ref{eq:q2_action}). It is sharply peaked at $\tfrac{m}{w-1}\approx \tfrac{n}{w}$. The weighting function $\hat{\delta}^{(w)}(m;n)$ in eq.~(\ref{eq:s2_action}) is instead
\be\label{deltahatmn}
\hat{\delta}^{(w)}(m;n) =
\begin{tikzpicture}[baseline=-.5ex]
    \node[](n1)at(0,0){$\bar{\alpha}^1_{1+\frac{m}{w+1}}$};
    \node[](n2)[right=-1ex of n1]{$\alpha^2_{-1-\frac{n}{w}}$};
    \draw[](n1.north)-- ([yshift=1ex]n1.north)--([yshift=1ex]n2.north)--(n2.north);
\end{tikzpicture}= \delta^{(w+1)}(n;m)^* \ .
\ee

\end{document}